\documentclass[reprint,amsmath,amssymb,aps,floatfix]{revtex4-2}
\usepackage[margin=0.75in]{geometry}
\usepackage[utf8]{inputenc}

\usepackage{amsmath}
\usepackage{enumitem}
\usepackage{amssymb,float}
\usepackage{soul}
\usepackage[normalem]{ulem}
\usepackage{amsfonts}
\usepackage{amstext}
\usepackage{mathtools}
\usepackage{amsthm}
\usepackage{subcaption}
\usepackage{graphicx}
\usepackage{color}
\usepackage{lipsum}
\usepackage{natbib}
\usepackage{xcolor}
\usepackage{float}
\usepackage{bm}
\usepackage{hyperref}
\usepackage{comment}
\usepackage[normalem]{ulem}
\usepackage{ragged2e}
\usepackage[font={scriptsize}, skip=0pt,format=plain,singlelinecheck=false]{caption}

\usepackage{helvet}
\def\be{\begin{equation}}
\def\ee{\end{equation}}
\def\bea{\begin{eqnarray}}
\def\eea{\end{eqnarray}}

\def\be{\begin{equation}}
\def\ee{\end{equation}}
\def\bea{\begin{eqnarray}}
\def\eea{\end{eqnarray}}

\begin{document}
\title{Ecosystems as adaptive living circuits}

\author{Ankit Dhanuka$^1$}  
\author{Avi I. Flamholz$^2$} 
\author{Arvind Murugan$^3$}
\email{amurugan@uchicago.edu}
\author{Akshit Goyal$^4$} 
\email{akshitg@icts.res.in}
\address{$^1$ National Centre for Biological Sciences-TIFR, Bengaluru 560065}
\address{$^1$ The Laboratory of Environmental Microbiology, The Rockefeller University, New York NY 10065}
\address{$^3$ Department of Physics, University of Chicago, Chicago, IL 60637}
\address{$^4$ International Centre for Theoretical Sciences, Tata Institute of Fundamental Research, Bengaluru 560089}

\begin{abstract}
\noindent Unlike many physical nonequilibrium systems, in biological systems, the coupling to external energy sources is not a fixed parameter but adaptively controlled by the system itself. 
We do not have theoretical frameworks that allow for such adaptability. As a result, we cannot understand emergent behavior in living systems where structure formation and non-equilibrium drive coevolve. 
Here, using ecosystems as a model of adaptive systems, we develop a framework of living circuits whose architecture changes adaptively with the energy dissipated in each circuit edge.  
We find that unlike traditional nonequilibrium systems, living circuits exhibit a phase transition from equilibrium death to a nonequilibrium dissipative state beyond a critical driving potential. This transition emerges through a feedback mechanism that saves the weakest edges by routing dissipation through them, even though the adaptive rule locally rewards the strongest dissipating edges. Despite lacking any global optimization principle, living circuits achieve near-maximal dissipation, with higher drive promoting more complex circuits. Our work establishes ecosystems as paradigmatic examples of living circuits whose structure and dissipation are tuned through local adaptive rules.
\end{abstract}

\maketitle
\noindent
Many living processes have been studied through the lens of non-equilibrium dynamics \cite{fang2019nonequilibrium}. Deviations from equilibrium enable numerous biological functions, including transport \cite{schmiedl2008efficiency,mugnai2020theoretical,albaugh2022simulating}, synthesis  \cite{andrieux2008nonequilibrium,nguyen2016design,arango2019self}, pattern formation \cite{zhang2023free,horowitz2017minimum,cross1993pattern}, and information processing tasks such as error correction \cite{hopfield1974kinetic,ninio1975kinetic,murugan2014discriminatory, owen2023size} and molecular computation \cite{mehta2012energetic,ouldridge2017thermodynamics,harvey2023universal,mahdavi2024flexibility,floyd2024limits}. Typically, these frameworks assume a fixed coupling to external energy sources that drive systems away from equilibrium \cite{qian2007phosphorylation,hill2005free}, focusing primarily on how adaptation or other functions depend on energy consumption (dissipation). The extent of dissipation itself is rarely considered as an adaptive, dynamic variable; rather, it is viewed as a fixed driver of adaptation in other degrees of freedom.

Yet, a critical aspect of living matter is that energy dissipation is not static but an adaptive dynamical degree of freedom, which may change as the system self-organizes into different states. For instance, within a cell, the cytoskeleton is continually reconfigured by ATP consumption by motor proteins but the resulting configuration can alter how much ATP is consumed \cite{mizuno2007nonequilibrium,sanchez2012spontaneous,sakamoto2024f}. On a larger length scale, consider a Winogradsky column, a classic model of a complex microbial ecosystem. A transparent cylinder containing well-mixed pond sediment is sealed and exposed to sunlight \cite{madigan1997brock,esteban2015temporal}. Although it lacks spatial structure at first, phototrophs soon capture sunlight and produce oxygen and organic byproducts, which drive the formation of spatially structured microbial layers. This emergent stratification enhances nutrient recycling \cite{madigan1997brock}, improving conditions for the phototrophs and leading to increasingly efficient solar energy capture over time. Similar structure-energy feedback loops occur globally, as ecosystems organize the planetary environment in ways that impact solar energy capture, thereby driving further structural reorganization. Such feedback loops between structure and energy dissipation can lead to distinct organizational states over longer timescales; these outcomes cannot be predicted using traditional frameworks that assume fixed energy couplings. Understanding such adaptive dissipation is crucial for identifying factors that promote stronger dissipation and complex organization versus those that drive systems toward equilibrium and eventual death.

\begin{figure*}[ht!]    
   \centering\includegraphics[width=\textwidth]{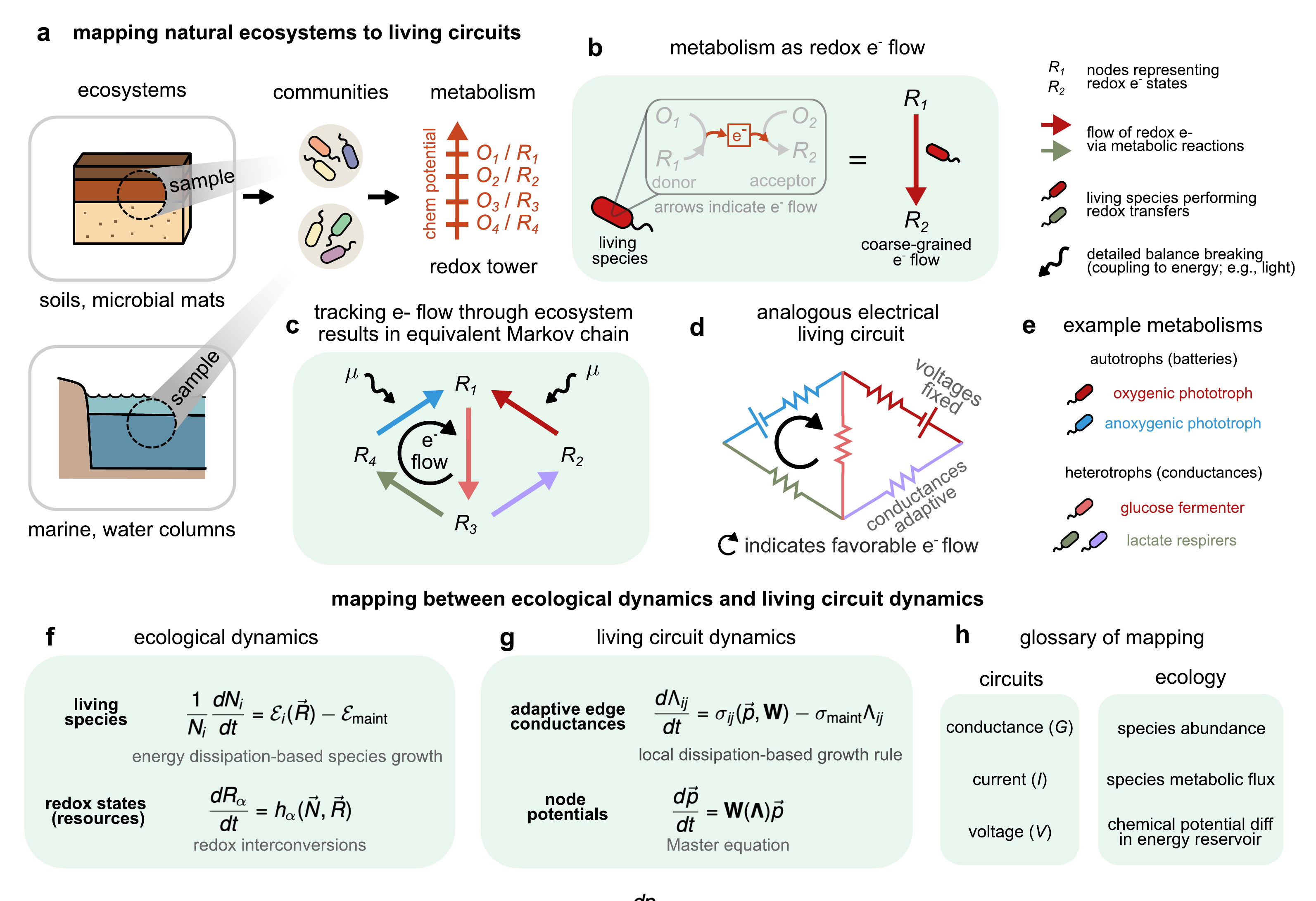}
    \caption{\justifying\textsf{\textbf{Exact mapping from ecology to adaptive living circuits.} (a) Microbial ecosystems sustain themselves on energy extracted through redox metabolism. We rigorously map any such ecosystem to an equivalent living circuit whose conductances adjust based on local rules driven by energy dissipation. (b) Through redox metabolism, living species (here, red microbe) transfer electrons from one redox state to another (here, $R_1 \to R_2$). 
    Shown is internal electron transfer within a living species, which can be coarse-grained to a directed edge tracking the e$^-$ state from $R_1$ to $R_2$ (states denoted by molecules).
    (c) Tracking transitions between redox states through the whole ecosystem results in a non-equilibrium Markov chain. Since the final redox state of one species is the initial state for another, e$^-$ create a web of metabolic interactions between species.  Each edge $ij$ of this Markov chain is an edge catalyzed by living species, as shown in (b), with direction indicating the net flow of e$^-$. Autotroph edges (with wiggled arrows on top) extract energy from a reservoir with driving potential $\mu$; they break detailed balance and promote the circuit out of equilibrium.
    (d--e) Analogous electrical circuit representation of the Markov chain. Living species represent conductances that dynamically tune themselves based on local dissipation. The coupling of some redox transformations to an energy source (e.g., light in phototrophs) serves as a battery. (f--g) Equations that map (f) ecological consumer-resource models and (g) living circuits. Ecological species growth dynamics based on energy dissipation motivate the dynamics of circuit edge conductances which increase with local dissipation along the edge. Ecological resource dynamics map to the Master equation for the circuit, with steady-state flux balance for electrons mapping to Kirchhoff's current law. (h) Glossary of circuit quantities mapped to their ecological analogs.}
}
\label{fig1}
\end{figure*}

Here, we introduce a simplified theoretical framework for structure-energy feedback, termed ``living circuits’’, using microbial ecosystems as a model. We demonstrate an exact mapping of the dynamics of microbial ecosystems onto electrical circuits whose link conductances change dynamically. We achieve this mapping by shifting focus from microbial species and nutrient flows to the mathematically equivalent yet simpler framework of electron flow through redox reactions. Extending Szent-Gyorgyi's principle of `following the electron' \cite{szent2012introduction} from single species to entire ecosystems, we find that species abundances, traditionally the focus of ecological models, play the role of conductances (Fig.~\ref{fig1}). By assuming that species growth rates are set by metabolic energy availability, ecosystem dynamics reduce to circuits whose conductances evolve according to local energy dissipation. Despite its simplicity, this local adaptation rule gives rise to complex global arrangements, as energy dissipation across each link depends intricately on the conductances of other network links. Consequently, we find that living circuits may evolve either towards highly dissipative states, corresponding to functional ecosystems with high energy capture or towards equilibrium and death in different parameter regimes. 
In this way, living circuits are a simple framework that capture the fundamental physics of complex adaptive systems that self-organize their displacement from equilibrium.

\noindent
\textbf{\textsf{Mapping redox metabolism in ecosystems to living circuits.}} Microbial ecosystems exist in diverse environments such as soil microbial mats, Winogradsky columns, water columns, and anaerobic digesters (Fig. \ref{fig1}a). While microbes interact through multiple mechanisms\cite{hibbing2010bacterial,mitri2013genotypic} including phage exchange \cite{soucy2015horizontal}, toxin secretion \cite{cornforth2015antibiotics}, and other signaling processes \cite{keller2006communication}, a fundamental driving force across these ecosystems is the exchange of redox resources needed for metabolism and growth\cite{falkowski2008microbial,delattre2019consistent,von1999does,von2006thermodynamics}. Microbial metabolism is based on redox reactions, where species gain energy by moving electrons from one redox state to another\cite{lehninger2005lehninger,heijnen1999bioenergetics,goyal2023closed}. This transfer of electrons occurs by coupling two half-reactions involving resources (Fig. \ref{fig1}b): one resource $R_i$ is oxidized to $O_i$, while the other $O_j$ is reduced to $R_j$. In the process, electrons move from one redox state to another, starting from a state where an electron is present on $R_i$ and ending at a state where it is on $R_j$. The change in redox chemical potential from one state to the other quantifies the energy gained during the electron transfer.

By focusing on terminal redox states in each species' metabolism (labeled $i$ to denote electrons present on reduced resource $R_i$), we can map an entire ecosystem to an electrical circuit---or equivalently, a Markov chain---tracking the flow of electrons through the ecosystem (Fig. \ref{fig1}c; Appendix~\ref{Methods}--\ref{mappingAppendix}). In this representation, nodes correspond to redox states $i$, while edges represent transitions between states catalyzed by living species. This network generalizes Szent-Gyorgyi's ``follow the electron'' \cite{szent2012introduction} approach to entire ecosystems (Fig.~\ref{mapping_met_network_app_fig}). The circuit or Markov chain contains $N$ nodes representing redox states $i$ whose probabilities $p_i$ are governed by a master equation \cite{schnakenberg1976network}:

\begin{equation}
\frac{dp_i}{dt} = \sum_{j=1}^N W_{ij} p_j
\end{equation}
where $W_{ij}$ describes the transition rate from state $j \rightarrow i$, which is set by $W_{ij} = \Lambda_{ij}k_{ij}$. Here, $\Lambda_{ij}=\Lambda_{ji}$ represents the abundance of species catalyzing this electron transfer, and $k_{ij}$ the kinetic rate of the $j \rightarrow i$ transition (Appendix~\ref{Methods}). For simplicity we focus on interconversions between redox states, where rates depend only on probabilities $p_i$ and not on the total resource amounts $O_i$ and $R_i$. This is equivalent to assuming that the microbial ecosystem is quasi-closed to material exchange on the timescale of the dynamics. Material exchange can easily be incorporated by adding sources and sinks to the Markov chain \cite{norris1998markov}.

Edges of the circuits correspond to living species, with edge conductances $\Lambda_{ij}$ being analogs of species abundances. These circuits are ``living'' since edge conductances $\Lambda_{ij}$ dynamically adapt, growing and dying based on local dissipation rules:

\begin{equation}
\frac{d\Lambda_{ij}}{dt} = \sigma_{ij}(\vec{p}, \textbf{W}) - \sigma_{\text{maint}}\Lambda_{ij}
\label{livingLamEq}
\end{equation}

where:

\begin{eqnarray}
    \sigma_{ij}(\vec{p}, \textbf{W}) &=& ( W_{ij} p_j - W_{ji} p_i) \log\left(\frac{W_{ij}p_j}{W_{ji}p_i}\right) \nonumber \\
    & =& \Lambda_{ij}( k_{ij} p_j - k_{ji} p_i) \log\left(\frac{k_{ij}p_j}{k_{ji}p_i}\right)
\end{eqnarray}
represents the \textit{total} energy dissipation along the $ij$ edge, and $\sigma_{\text{maint}}$ is the \textit{per capita} dissipation required for maintenance. This is analogous to ecological species growth dynamics based on energy acquisition \cite{goyal2023closed} (Fig. \ref{fig1}f).

Special living species---autotrophs---break detailed balance in this circuit by coupling some of these transformations $i \to j$ to external energy sources. For example, phototrophs couple redox transformations to light, while chemolithoautotrophs couple to favorable e$^-$ donors such as Fe$^{2+}$ \cite{madigan1997brock}. We represent this coupling mathematically through modified transition rates: $k_{ij}/k_{ji} = (\tilde{k}_{ij}/\tilde{k}_{ji})e^\mu$, where $\mu$ quantifies the chemical potential or nonequilibrium driving potential provided by photons or other sources of chemical energy, and $\tilde{k}$ represents rates absent external energy. This coupling sustains nonequilibrium dynamics in the circuit and completes our mapping from closed ecosystems to living circuits.

To parametrize real microbial ecosystems as circuits, we infer $k_{ij}$ from chemical potentials on the redox tower using known  thermodynamic relationships (Appendix~\ref{redoxTowerAppendix}). We find that abstract circuits where we sample $k_{ij}$ as random parameters (Appendix~\ref{Methods}) show the same phenomenology (Fig.~\ref{real_redox_tower}). While this parametrization does not affect our core findings, it renders the model analytically tractable (Appendices~\ref{critmuappendix}--\ref{satur_complexity}). Hence, in the rest of the manuscript we will use use random parameters to study circuit phenomenology.

\vskip 10pt
\noindent
\textbf{\textsf{Nonequilibrium steady states in living circuits collapse below a threshold detailed balance breaking.}} 
Physical systems driven away from equilibrium typically maintain nonequilibrium steady states (NESS) whenever external driving forces break detailed balance. In such systems, the emergence of a NESS is guaranteed for any nonzero driving potential \cite{qian2007phosphorylation,hill2005free}. Our living circuits, however, exhibit a fundamentally different behavior since they are self-organized through local adaptive rules.

\begin{figure*}
\centering\includegraphics[width=1.\linewidth]{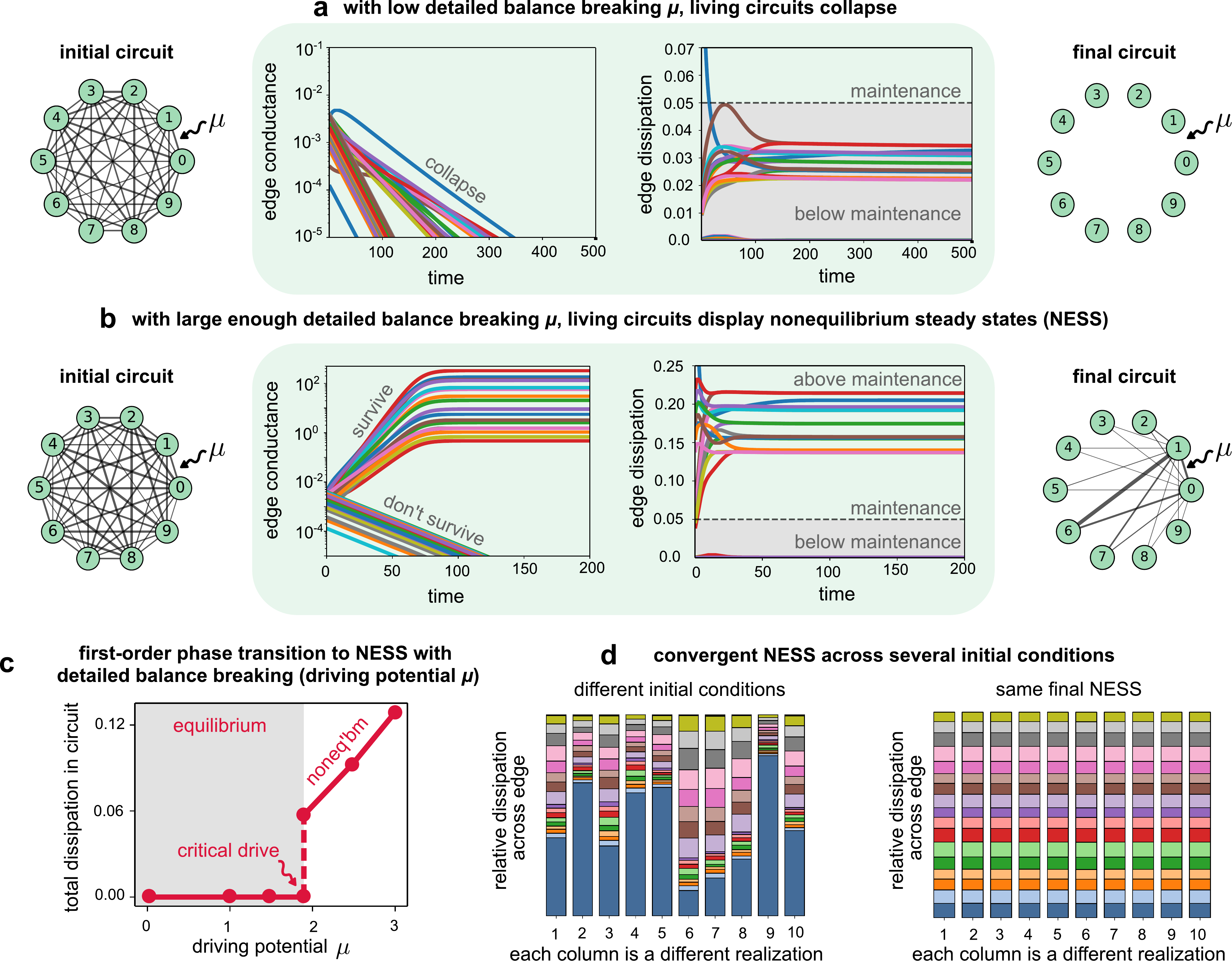}
\vspace{0.5cm}
    \caption{\justifying \textsf{\textbf{Non-equilibrium steady states (NESS) in living circuits collapse below a threshold detailed balance breaking.} 
(a) An initially complex circuit with 10 nodes (redox states) and 45 edges (living species) fails to reach a NESS despite a driving potential $\mu=1$ that breaks detailed balance (Appendix~\ref{Methods}); this is analogous to collapse of an ecosystem with insufficient energy. Collapse occurs because of dynamic edge conductances, which shrink over time due to insufficient dissipation. Plots show dynamics of edge conductances and dissipation. Undirected edges represent species which catalyze both forward and reverse transitions between nodes.
(b) With sufficient energy drive $\mu=3$, the same initial circuit self-organizes to a NESS; the final adapted circuit has a different topology after losing some edges (extinct species) and strengthening others during dynamics, with surviving edges dissipating more than $\sigma_{\text{maint}}$. Edge thicknesses are proportional to edge conductances.
(c) Plot of average global dissipation in living circuits with drive $\mu$ showing that NESS self-organize only above a critical drive $\mu_{\text{crit}} \approx 2$; this is in contrast with conventional non-equilibrium circuits where NESS are instead predestined, i.e., have nonzero dissipation for any nonzero drive.
(d) Stacked bar plots showing convergence of different initial circuits, with the same redox networks, to the same final NESS. Each column represents one initial condition; colors indicate different edges, with heights indicating normalized edge dissipations.}}
\label{fig2}
\end{figure*}

To investigate the self-organized nature of NESS, we simulate the behavior of living circuits with varying driving potentials $\mu$. We initialize a complex circuit with 10 nodes (redox states) and 45 edges (living species), and analyze its dynamics under different driving potentials. At low drive ($\mu = 1$), the circuit fails to reach a non-trivial NESS despite the presence of an external energy source that breaks detailed balance (Fig. \ref{fig2}a). This failure manifests as a complete collapse of the network structure, with edge conductances decaying over time due to insufficient dissipation. 
In contrast, when the same initial circuit is subjected to sufficient drive ($\mu = 3$), it self-organizes to a robust NESS (Fig. \ref{fig2}b). During this process, the circuit undergoes significant topological changes: some edges disappear (extinct species) while others strengthen, resulting in a network with fewer active edges that each dissipate significantly more than maintenance dissipation $\sigma_{\text{maint}}$.

\begin{figure}
   \centering\includegraphics[width=\linewidth]{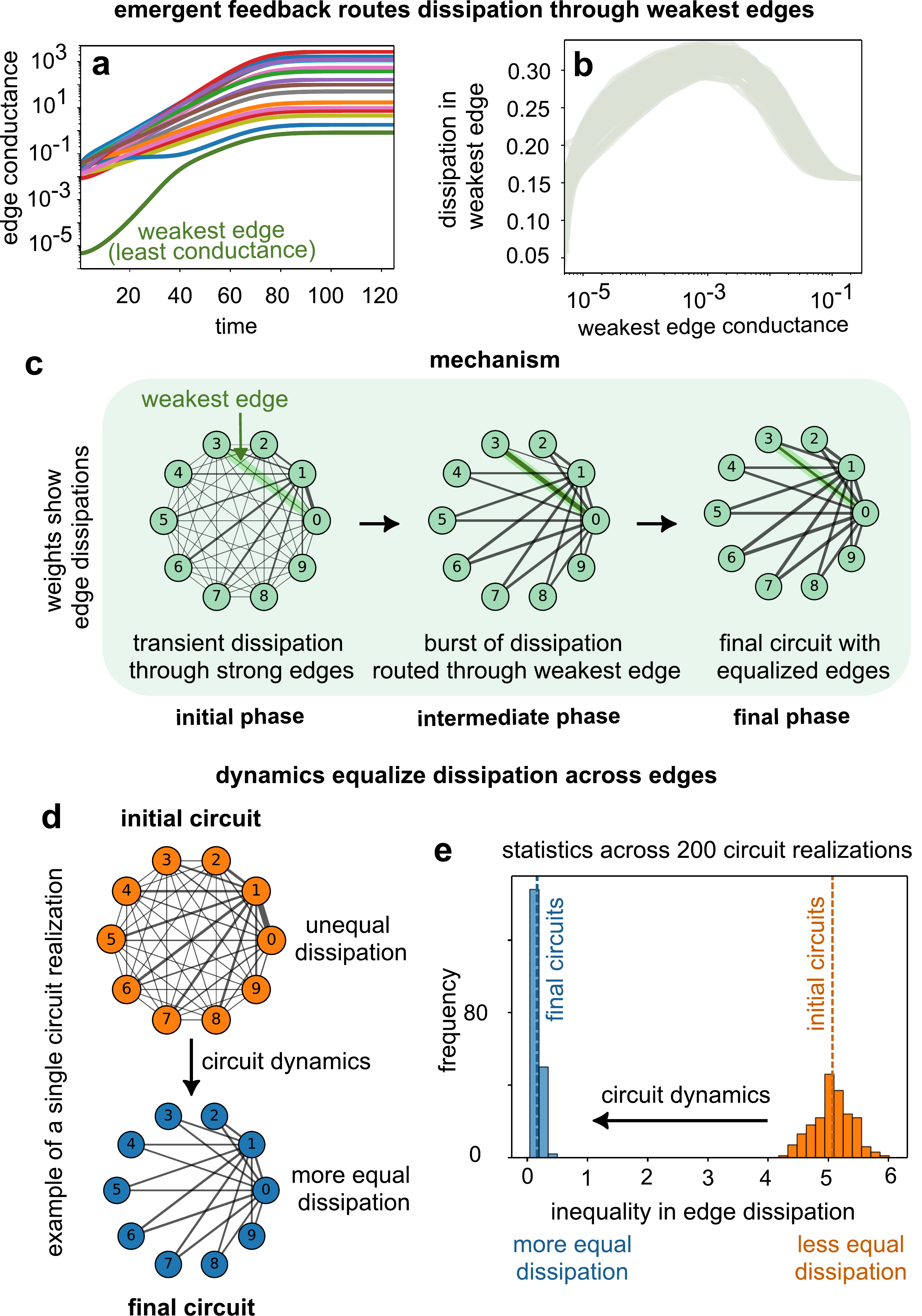}
     \vspace{0.1cm}
      \caption{\justifying \textsf{\textbf{Emergent robustness during adaptation stabilizes living circuits.}
 (a) Conductance dynamics in a living circuit starting with the weakest edge having nearly zero conductance (green); as the circuit approaches NESS the weakest edge is rescued by routing more dissipation through it and thus increasing its conductance. (b) Plot of local dissipation along the weakest edge as a function of its conductance during dynamics; dissipation is non-monotonic, increasing when the edge conductance is low and decreasing once the conductance becomes sufficiently large. (c) Mechanism behind non-monotonicity and ``save the weakest'' effect: the initial increase in dissipation occurs because while the edge is weak, the rest of the circuit sets the density of nodes (electron states), often misaligned with the kinetics ($k_{ij}$) of the weakest edge. Thus, edges near collapse typically have highest dissipation, leading to a burst of dissipation through the edge, increasing its conductance and rescuing it. Eventually the circuit adjusts and prevents the edge from a runaway increase in conductance, stabilizing the circuit at the NESS.
(d) Example of a complex circuit initialized with unequal dissipation across edges (orange; uneven thicknesses); after dynamics to NESS, the final circuit has edges with more equal dissipation (even thicknesses). (e) Distribution of inverse evenness of edge dissipation across many realizations of circuits with distinct redox networks. Inverse evenness is measured as the coefficient of variation of edge dissipations in a circuit; final variability in blue is $\approx 10$x lower than initial (orange).}}
\label{fig3}
\end{figure}

Strikingly, we find that the network self-organizes into a non-trivial NESS only above a critical driving potential $\mu_{\text{crit}} \approx 2$ (Fig. \ref{fig2}c).
Unlike conventional nonequilibrium circuits, the total dissipation in living circuits shows a first-order phase transition. Below $\mu_{\text{crit}}$, there is no dissipation because circuits collapse. Right above it, dissipation jumps discontinuously to a value above $\sigma_{\text{maint}}$ (see Appendix \ref{1storderappendix}). Beyond $\mu_{\text{crit}}$, dissipation increases linearly with driving potential, as circuits adopt increasingly dissipative configurations. We can analytically compute a relationship connecting the critical driving potential $\mu_{\text{crit}}$ with the ratio of $\sigma_{\text{maint}}$ and an emergent current $\chi(\mathbf{k})$ which is set by the chemistry 
(see Appendix \ref{critmuappendix}),  

\begin{equation}
    \left(e^{\mu_{\text{crit}}} - 1\right) \cdot \mu_{\text{crit}} = \frac{\sigma_{\text{maint}}}{\chi(\mathbf{k})}
\end{equation}

The existence of a critical drive is non-trivial and captures the essence of structure-energy feedback. In principle, edge dissipation could increase arbitrarily by scaling up conductances $\mathbf{\Lambda}$. However, the resulting electron redistribution through the circuit creates a feedback loop that, below $\mu_{\text{crit}}$, invariably reduces per-capita dissipation $\sigma_{ij}/\Lambda_{ij}$ below $\sigma_{\text{maint}}$, causing edge extinction. We find that this feedback is independent of edge conductances $\mathbf{\Lambda}$ and instead depends strongly on $\mu$ and $\chi$. $\mu_{\text{crit}}$ is thus the drive at which circuit-wide feedback results in just enough dissipation to each edge to meet maintenance requirements.
Despite a first-order transition, different initial conditions with the same underlying redox network topology converge to identical final NESS (Fig. \ref{fig2}d), rather than showing hysteresis. This convergence indicates that the self-organized state represents a robust attractor in the space of possible network configurations, determined by the underlying redox chemistry rather than initial conditions. 


\vskip 10pt
\noindent
\textbf{\textsf{Emergent robustness during adaptation stabilizes living circuits.}}
Living circuits exhibit an emergent robustness during their adaptation: they can stabilize their weakest components through a cooperative feedback mechanism. As a consequence, living circuits are able to heal themselves and recover from energy-poor, near-death states.


We investigated this phenomenon by examining a living circuit initialized with a highly uneven distribution of edge conductances. We specifically focused on the dynamics of the weakest edge—corresponding to a biological species on the brink of extinction—with an initial conductance nearly $10^3$ times smaller than the strongest edges in the circuit. 

In conventional circuits, an edge with $10^3$ times lower conductance than the rest of the network will have negligible impact on system dynamics and stay that way. However, in living circuits, the conductance of this weakest edge increases dramatically during the approach to steady state, eventually stabilizing at a value comparable to other edges (Fig. \ref{fig3}a). 

The mechanism behind this ``save the weakest'' effect emerges from a non-monotonic relationship between edge conductance $\Lambda_{ij}$ and local dissipation $\sigma_{ij}$. We can calculate the conductance dynamics for the weakest edge (when $\Lambda_{\text{weak}} \ll \langle\Lambda_{ij}\rangle$) and find that it approximately follows a quadratic potential as (see Appendix \ref{xlogxappendix}):

\begin{equation}
\frac{d\Lambda_{\text{weak}}}{dt} \approx \frac{d}{d\Lambda_{\text{weak}}} \left(\frac{1}{2} \Xi_{\text{eff}} \Lambda_{\text{weak}}^{2}\right)
\end{equation}
where the effective curvature $\Xi_{\text{eff}}$ depends on circuit parameters. Interestingly, we find that the sign of this curvature changes at the phase transition: $\Xi_{\text{eff}}<0$ if $\mu<\mu_{\text{crit}}$ and $\Xi_{\text{eff}}>0$ if $\mu > \mu_{\text{crit}}$. Hence, with insufficient drive, the weakest edge collapses as all edges should. With above-critical drive, $\Lambda_{\text{weak}}$ always grows when the edge is weak. This approximation eventually breaks down as the conductance grows stronger. As shown in Fig. \ref{fig3}b, simulations confirm that the weakest edge dissipation behaves non-monotonically with its conductance. This counterintuitive effect can be understood through the circuit's collective dynamics: when an edge is weak, the rest of the circuit determines the electron distribution across nodes, often misaligned with the kinetics ($k_{ij}$) of the weakest edge. This misalignment creates a transient electron flow through other edges in the circuit (Fig. \ref{fig3}c), resulting in a burst of dissipation through the weakest edge when its conductance is low. The increased dissipation drives conductance growth, effectively rescuing edges near collapse. Eventually, as the edge strengthens, the circuit adjusts to prevent runaway growth in conductance, stabilizing at the NESS.

To quantify this stabilization effect, we examined how the distribution of edge dissipations evolves during circuit dynamics. Starting from initial configurations with highly uneven distributions (Fig. \ref{fig3}d,e, orange distribution with high coefficient of variation), living circuits self-organize to configurations where dissipation is more uniformly distributed across edges (Fig. \ref{fig3}d,e, blue distribution). This equalization of dissipation is a hallmark of the adaptive nature of living circuits, where edges collectively adjust to maintain system-wide stability.

The equalization process can be quantified by measuring the coefficient of variation of edge dissipations, which decreases approximately tenfold during self-organization. This demonstrates that edges with initially disparate dissipations converge to a more uniform distribution at steady state, effectively preventing extinction of weaker components. Unlike conventional circuits, where weak components remain weak or fail entirely, living circuits actively redistribute energy to maintain diverse edge configurations.

This emergent stabilization does not seem to be a feature of other adaptive systems such as vasculature or \textit{Physarum}. There, pruning weak links is often economical—minimizing cost by eliminating low-conductance edges that contribute little \cite{ronellenfitsch2016global,alim2013random}. In our framework, each link represents a selfish agent that pays its own cost via local energy acquisition; there is no global resource pool to conserve by removing weak components. As a result, species near extinction may be rescued by community-level feedbacks that transiently increase their per-capita dissipation when their abundances are low. This leads to more uniform energy allocation across edges and buffers communities against fluctuations in species abundances. It also renders ecosystems more stable against external perturbations, such as sudden or time-varying changes in driving potential (Appendix \ref{timeVarEnvAppendix}). Finally, saving the weakest benefits the entire system by allowing all species to extract more energy from the environment.


\begin{figure*}
     \includegraphics[width=0.6\linewidth]{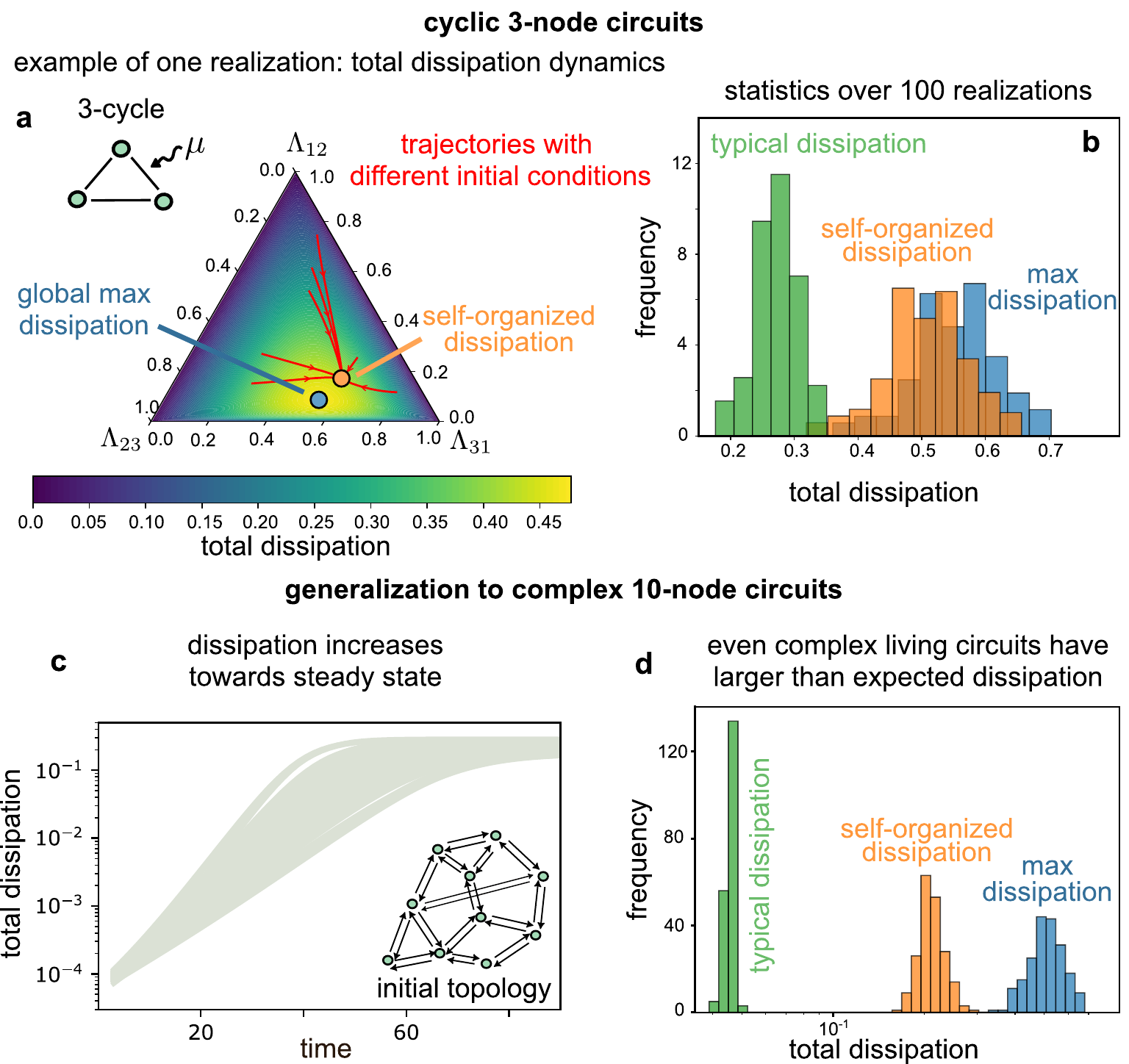}
     \vspace{0.1cm}
      \caption{\justifying \textsf{\textbf{Living circuits dissipate near-optimally despite no global optimization principle or Lyapunov function.} 
(a) Landscape of global dissipation in a simple circuit with 3 nodes and 3 edges. Simplex of normalized conductances of all 3 edges, overlayed with a heatmap of total dissipation. All initial conditions evolve to the same living circuit steady state (red point), whose dissipation is close to the global maximum (blue point) for the chosen redox kinetics $k_{ij}$ and driving potential $\mu$.
(b) Histogram of global dissipation over many realizations of $k_{ij}$ and $\mu$, with living circuit steady state dissipation (orange), global maximum dissipation (blue) and typical dissipation (green), indicating the dissipation expected for randomly chosen fixed conductances. Living circuit steady state dissipation is nearly-optimal, being much closer to the max than expected.
(c) Near-optimal dissipation generalizes to more complex circuits (here with 10 nodes and 45 edges; both forward and reverse transitions shown);
Dynamics of global circuit dissipation for 100 realizations of random complex circuits, showing that dissipation generally increases over evolution to a living circuit steady state. 
(d) Histogram of global dissipation, similar to (b) for complex 10-node circuits, showing that near-optimal dissipation is a robust phenomenon.}}
\label{fig4}
\end{figure*}

\vskip 10pt
\noindent
\textbf{\textsf{Living circuits dissipate near-optimally despite no global optimization.}} In our living circuit model, the conductance of each edge grows `greedily', with more dissipative edges attracting more flux and thus dissipating more. 
As a result, the dynamics of a living circuit, when the adaptive dynamics (Eq. (\ref{livingLamEq})) for $\Lambda_{ij}$ are included, lack any obvious global optimization principles such as a Lyapunov function (see Appendix \ref{gradDesAppendix}). 
Despite this absence of global optimization, we find that living circuits self-organize to a nonequilibrium steady state (Fig. \ref{fig4}a, red dot) whose global dissipation nearly reaches the theoretical maximum (Fig. \ref{fig4}a, blue dot). This global maximum represents the set of edge conductances that maximize total dissipation for a given redox network with fixed microscopic kinetics $k_{ij}$.

Fig. \ref{fig4}a demonstrates this phenomenon for a specific redox tower with 3 nodes (redox states) and 3 edges (living species), where all circuit configurations can be visualized on a simplex of normalized conductances. The heatmap overlay shows the landscape of global dissipation as a function of edge conductances. Strikingly, all initial conditions self-organize to the same steady state with near-maximal dissipation. Across several realizations of such simple 3-node circuits, the self-organized dissipation (Fig. \ref{fig4}b, orange) is much closer to the global maximum (Fig. \ref{fig4}b, blue) than expected by chance (Fig. \ref{fig4}b, green). For cycles with $N$ nodes, we can analytically calculate the relationship between edge adaptation dynamics and total dissipation $\sigma_{\text{tot}}$ in living circuits and find that they nearly maximize total (or global) dissipation (see Appendix \ref{gradDesAppendix}):

\begin{equation}
    \frac{d\Lambda_{ij}}{dt} \approx f\Big(\frac{1}{\mathbf{\Lambda}}\Big) \cdot \frac{d\left( \log\sigma_{\text{tot}}\right)}{d\Lambda_{ij}} -  \sigma_{\text{maint}}\Lambda_{ij},
\end{equation}
where $f\Big(\frac{1}{\mathbf{\Lambda}}\Big)$ is a positive-valued function of all edge conductances with parameter dependence on circuit chemistry. The first term amounts to an emergent desire to maximize the total dissipation $\sigma_{\text{tot}}$, even though the actual postulated dynamics (Eq. (\ref{livingLamEq})) are entirely selfish and unaware of $\sigma_{\text{tot}}$. On the other hand, the second term with $\sigma_{\text{maint}}\neq0$ is a correction that shows that circuit dynamics deviate from maximizing total dissipation, but may instead dissipate near-optimally. To confirm that near-optimal dissipation is a robust phenomenon beyond cyclic topologies, we extended our analysis to more complex circuits with 10 nodes and 45 possible edges (Fig. \ref{fig4}c). Unlike 3-node circuits with only one self-reinforcing cycle, these circuits have much more complex topologies with many interlocking cycles that may compete to lower overall dissipation. Despite this complexity, the global dissipation of these circuits typically increases during adaptation (Fig. \ref{fig4}c). Consequently, their final self-organized dissipation (Fig. \ref{fig4}d, orange) remains $\approx 7$x closer to the global maximum (Fig. \ref{fig4}d, blue) than expected by chance (Fig. \ref{fig4}d, green). 

\begin{figure}[t!]
    \hspace{-1.1cm}\includegraphics[width=1.1\linewidth]{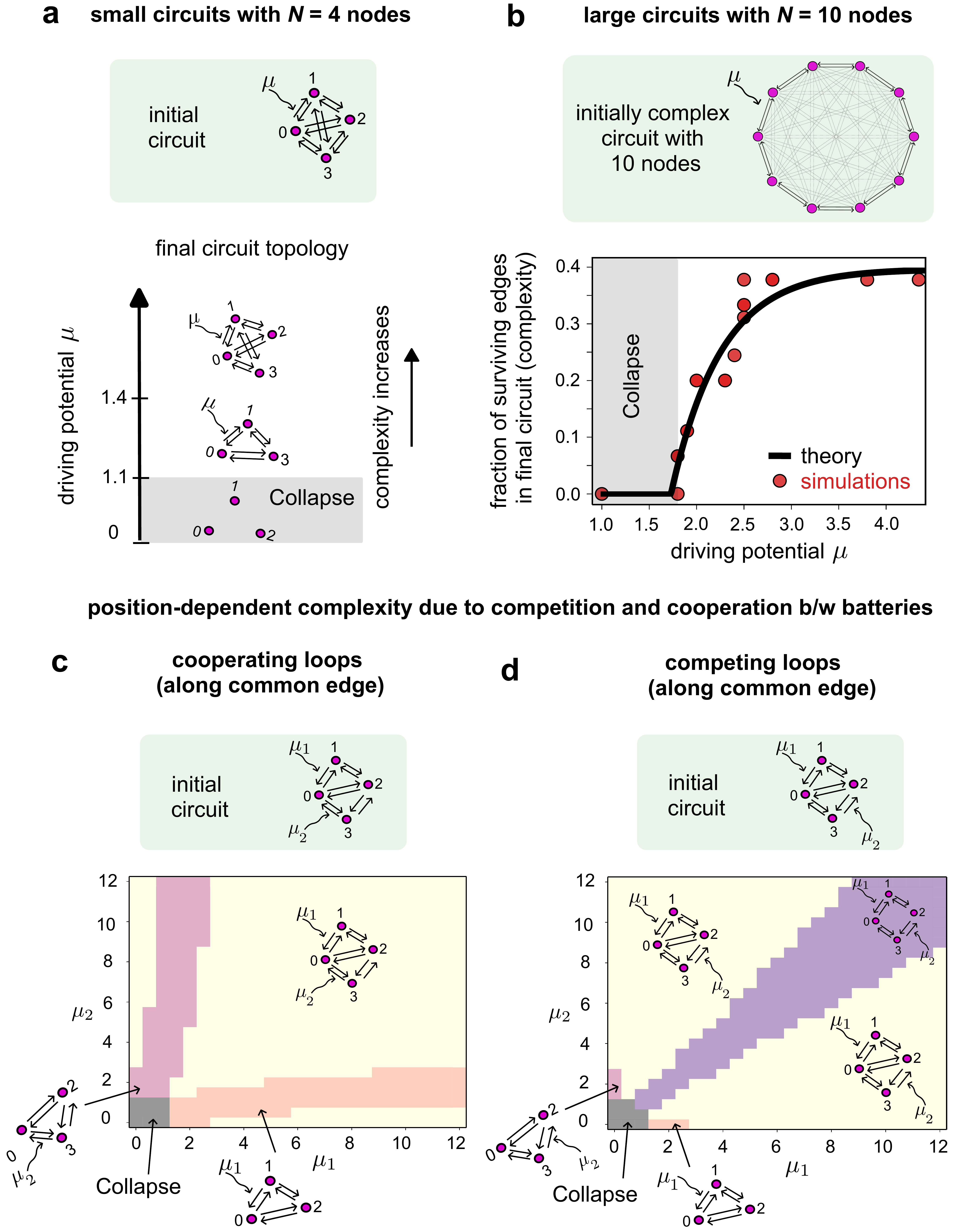}
        \vspace{0.1cm}
      \caption{\justifying \textsf{\textbf{Increasing nonequilibrum drive promotes circuit complexity.}
(a) Topology of self-organized circuits with increasing drive ($N=4$ nodes); circuits become more complex (with more edges) with increasing $\mu$. (b) Plot of circuit complexity with increasing energy drive for a complex circuit with 10 nodes. Higher drive promotes complexity; at large enough drive, complexity saturates. Points show simulations; line shows 1-parameter fit to theoretical predictions of critical drive $\mu_{\text{crit}}$ and saturating complexity $4/N$ (Appendix \ref{critmuappendix} and \ref{satur_complexity}). 
(c, d) Phase diagram of circuit topology with two drives (batteries), representing ecosystems with two autotrophs coupling to energy sources with drives $\mu_1$ and $\mu_2$; initial circuits share a common edge. Colors represent distinct circuit topologies, as shown. (c) When autotrophs cooperate along the common edge (dissipation from both adds up), increasing drive evenly ($\mu_1 \approx \mu_2$) promotes the most complex circuit (yellow). (d) When autotrophs compete instead (dissipation subtracts along common edge), the most complex circuit requires uneven energy drive (yellow vs. purple). Cooperation and competition depends on relative position of autotrophs; in random networks autotrophs typically cooperate (Fig. \ref{fig:multBattFig}).
}}
\label{fig5}
\end{figure}

These results suggest that a ``near-maximum dissipation principle'' might hold for general biological systems that self-organize through local energy-dependent growth rules. 
Adaptive living systems have long been described as minimizing dissipation, consistent with an efficient distribution of resources. Examples include flow networks in slime moulds and vascular networks in circulatory systems \cite{hu2013adaptation,ronellenfitsch2016global,corson2010fluctuations}.
In contrast, for some far-from-equilibrium systems --- including living systems --- it has been hypothesized that these systems may organize to maximize dissipation or entropy production \cite{Martyushev2006MEP,Juretic2003PhotosynthesisMEP,Paltridge1979ClimateDissipation}, though this has not been shown to arise from first principles. 
Our work shows that living circuits are a class of nonequilibrium systems that may achieve near-maximum dissipation through the collective interaction of components each following simple local rules.

\vskip 10pt
\noindent
\textbf{\textsf{Nonequilibrium drive promotes circuit complexity.}} The emergence of complex self-organized structures is a hallmark of living systems \cite{prigogine1977self, england2013statistical,england2015dissipative}. Do nonequilibrium systems inherently trend toward greater complexity, as suggested by theories of dissipative structures \cite{prigogine1970self,prigogine1977self} and complexity ratchets \cite{perunov2016statistical}? We can directly address this puzzle in our living circuits; we do so by analyzing how the driving potential $\mu$ affects circuit complexity, defined as the fraction of edges (living species) that survive in the self-organized NESS.

We first examined how increasing the driving potential $\mu$ affects the topology of self-organized circuits with 4 nodes (redox states) and initially fully connected edges (Fig. \ref{fig5}a).  At low driving potential ($\mu < 1.1$), the final circuit collapses entirely, with all edges falling below maintenance $\sigma_{\text{maint}}$. With slightly higher drive ($\mu = 1.1$), a simple 3-node cycle emerges, forming the minimal structure capable of maintaining a nonequilibrium current. Further increasing the drive ($\mu = 1.4$) leads to more complex topologies with additional edges, culminating in nearly fully connected networks at the highest energy drives.

This positive drive-complexity relationship generalizes to larger networks. For a 10-node redox circuit (Fig. \ref{fig5}b), we quantified complexity as the fraction of surviving edges after self-organization. Above the critical driving potential $\mu_{\text{crit}}$, complexity increases monotonically with driving force before eventually saturating for large drive. Notably, large circuits with $N=10$ nodes saturate at a complexity of $\approx 0.4$, far from full complexity. This saturation reflects a fundamental limit on the complexity of large circuits for a given set of redox kinetics $k_{ij}$. In the limit of low variance in kinetic coefficents, we can analytically compute the saturating complexity using graph-theoretic arguments as $\approx \frac{4}{N}$, which agrees well with simulations across a range of $N$ (Appendix \ref{satur_complexity}).

This limit to complexity can be somewhat overcome by multiple sources of nonequilibrium drive (batteries). This situation corresponds to having multiple autotrophs, such as oxygenic and anoxygenic phototrophs, that can coexist in the same environment. We find that multiple batteries makes the drive-complexity relationship in living circuits richer, allowing for greater complexity when drives are large (Fig. \ref{fig:multBattFig}). To explore this, we examined circuits with two light-coupled edges representing different autotrophs, each coupling to a distinct driving force with potentials $\mu_1$ and $\mu_2$ (Fig. \ref{fig5}c-d). We find that the relative positioning of these autotrophs critically determines whether they cooperate or compete. When autotrophs are positioned to cooperate along a common edge (Fig. \ref{fig5}c), where dissipation from both adds constructively, the most complex circuit topologies emerge when both drives are balanced ($\mu_1 \approx \mu_2$, yellow region). In contrast, when autotrophs are positioned to compete (Fig. \ref{fig5}d), where dissipation along the common edge partially cancels, complexity is maximized with asymmetric drives (yellow vs. purple regions). In randomly generated networks, we find that cooperation between autotrophs is more common than competition (Fig. \ref{fig:multBattFig}). These results demonstrate that nonequilibrium drive controls not only whether a living circuit can self-organize to a nonequilibrium steady state, but also determines the complexity of the resulting network structure. 

\vskip 10pt
\noindent
\textbf{\textsf{Discussion.}} Inspired by the need to understand how living systems adaptively control their coupling to external energy sources—a capacity absent in most physical nonequilibrium systems—we have introduced living circuits as a theoretical framework where systems self-organize to determine their own energy dissipation through local adaptive rules. This represents a departure from traditional nonequilibrium physics, where the distance from equilibrium is externally imposed rather than dynamically determined by the system itself.

A key feature of living circuits is a counterintuitive “save the weakest” effect: species on the brink of extinction can be often rescued by the dynamics of the whole circuit. This arises despite each species locally acting in a greedy, “rich-get-richer” fashion—edges that dissipate more energy grow faster. Because of circuit-level feedbacks, weak links can temporarily experience higher per-capita dissipation due to a mismatch between their kinetic rates and the node potentials set by the rest of the network. This triggers conductance growth and stabilizes these weak components. 

Seemingly similar adaptive rules have been proposed in other contexts, including vascular networks and slime molds, but they often differ in critical ways in their local form with global consequences. For instance, postulated rules often have opposite sign \cite{hu2013adaptation,ronellenfitsch2016global,alim2013random,corson2010fluctuations,tero2010rules}, i.e., pruning rather than reinforcing links with high dissipation.  Further, some of these variants tune conductance of links based on current in that link \cite{ronellenfitsch2016global,alim2013random,corson2010fluctuations,tero2010rules} rather than dissipation. 

Our work also differs in terms of global behavior from these models. Our local rules lead to maximizing global dissipation while these prior works report minimized dissipation. More importantly, instead of our `save the weakest' result, these works often report a `kill the weakest' effect in which the edges with the weakest current or dissipation are pruned away completely. This emergent kill-the-weakest effect is nevertheless seen as a feature in those systems because, e.g., \textit{Physarum} as a whole benefits from not having to pay the metabolic cost of maintaining a vein that was not contributing to transport anyway.  In contrast, in an ecosystem, maintenance of each edge does not come at a metabolic cost to a central organizer or entity. Thus, our `save the weakest' effect can be seen as a feature for ecosystems since it shows how locally selfish agents can nevertheless maintain states of high diversity.

More broadly, these distinctions can be seen as arising due to how natural selection over evolutionary time interacts with these adaptive rules. 
In physical networks \cite{dillavou2022demonstration,dillavou2022demonstration,stern2023physical} like \textit{Physarum} or vascular systems \cite{katifori2010damage,corson2010fluctuations,hu2013adaptation}, natural selection acts globally on the function of the whole network.  As a consequence, the local rules themselves are under evolutionary pressure precisely because of their global effects. Thus, local rules can evolve (through Darwinian evolution) to explicitly optimize some global objective (such as minimizing dissipation), or pruning weaker connections because of the global objective translates into a fitness cost for the host organism of these networks. 
In contrast, in ecosystems, the units of selection are the links (representing species) and not the whole network since Darwinian evolution does not act at the level of ecosystems\cite{lewontin1970units,lotka1922contribution}. Thus, the global properties we report here, such as maximal dissipation and the `save the weakest' effect, are emergent consequences of independently acting selfish local rules rather than a Darwinian-selected outcome. 

Related ideas of maximizing dissipation have been explored in statistical physics. The Maximum Entropy Production Principle (MEPP) has been proposed as a governing principle for far-from-equilibrium systems \cite{kleidon2010non,perunov2016statistical,Martyushev2006MEP,dewar2005maximum,juretic2003photosynthetic,odum1968energy,odum1973energy,lotka1922contribution}.  However, its mechanistic basis in many contexts has remained elusive. Our framework provides a mechanistic basis for a global MEPP-like principle by deriving it from local, biologically plausible rules. This aligns with recent work on dissipative adaptation in biological matter \cite{england2015dissipative,perunov2016statistical,england2013statistical} and externally-driven physical systems \cite{vzupanovic2004kirchhoff}.

There are several natural directions to extend this framework. First, incorporating spatial structure—such as diffusive transport or physical gradients—would enable studies of spatial self-organization, niche partitioning, and metabolic stratification as in Winogradsky columns \cite{babcsanyi2017biogeochemical,bush2017oxic,van1993microbial,chuang2025bacterial,goyal2025energy}. Second, in ecosystems where a substantial fraction of resource flux is diverted into biomass rather than energy dissipation, including flux-based local growth rules and more realistic redox chemistry, may be more appropriate. It would be interesting to see if global quantities like dissipation are still maximized in this context. Third, relaxing the assumption of a closed system by including sources and sinks at specific redox states would allow modeling of ecosystems embedded in larger environments. Each of these variants would expand the class of ecosystems to which our living circuits framework applies.

Although our living circuits framework was developed using microbial ecosystems as a model, it is easy to generalize it to other biological systems with feedbacks between structure and energy use, including neural networks with plastic synapses \cite{harris2012synaptic,li2020energy} and adaptive immune networks \cite{mayer2015well}. The specific adaptive rule used here—dissipation-driven growth—could be replaced or augmented by other biologically motivated rules, such as those that depend on local interactions, voltages, or currents, among others. Such extensions would reveal and capture a wider class of adaptive behaviors and provide a unified framework for studying nonequilibrium systems with structure-energy feedback. 

\vskip 10pt
\noindent
\textbf{\textsf{Acknowledgements.}} We thank K. Alim, E. Katifori, D. Lacoste, O. Rivoire, G. Salmon, M. Sireci, and S. Vaikuntanathan for valuable discussions. A.D. acknowledges support from NCBS-TIFR through a Simons-NCBS Career Development Fellowship. A.I.F., A.M., and A.G. acknowledge financial support from the NSF (PHY-1748958 to the Kavli Institute for Theoretical Physics).  A.I.F. acknowledges support from the Burroughs Wellcome Fund, the Irma T. Hirschl Trust, and the Robertson Foundation. A.M. acknowledges support from NIGMS of the National Institutes of Health under award number R35GM151211, National Science Foundation through the Center for Living Systems (grant no. 2317138) and DMR-2239801. A.G. acknowledges support from the Ashok and Gita Vaish Junior Researcher Award, the DST-SERB Ramanujan Fellowship, as well the DAE, Govt. of India, under project no. RTI4001.

\bibliography{ref}
\bibliographystyle{unsrt}
\vspace*{\fill}
\vspace*{\fill}
\makeatletter
\renewcommand{\thefigure}{S\@arabic\c@figure}
\makeatother
\setcounter{figure}{0}
\clearpage
\newpage
\onecolumngrid
\appendix
\makeatletter
\def\@seccntformat#1{\large\csname the#1\endcsname\quad}
\renewcommand\section{\@startsection{section}{1}{\z@}%
                                   {-3.5ex \@plus -1ex \@minus -.2ex}%
                                   {2.3ex \@plus.2ex}%
                                   {\normalfont\large\bfseries\centering}}
\makeatother
\begin{center}
\large\textbf{Appendices and Supplementary Material\\}
\Large\textbf{Ecosystems as adaptive living circuits}
\end{center}
\vskip 40pt
\section{Methods}\label{Methods}

All results presented in this paper were obtained through extensive numerical simulations performed in \textsf{Python}, utilizing the \texttt{solve\_ivp} function from the \textsf{SciPy} library for integrating systems of differential equations. The simulation framework was designed to capture the essential dynamical features of the system through the following key components:

\begin{enumerate}[leftmargin=*, label=(\arabic*)]
    \item \textbf{Detailed-balance non-adaptive transition matrix}: A baseline transition matrix satisfying detailed balance was constructed to represent static interactions among components of the network. The detailed-balance condition imposes that, for any closed loop in the network, the product of transition rates in the clockwise direction must equal the product of rates in the counter-clockwise direction. Formally, for a loop 
$i_1 \to i_2 \to i_3 \to \cdots \to i_k \to i_1$, this condition reads:
\begin{equation}
\frac{k_{i_1, i_2}k_{i_2, i_3} \cdots k_{i_{k-1}, i_k}}{k_{i_2, i_1}k_{i_3, i_2} \cdots k_{i_k, i_{k-1}}} = 1 \, .
\end{equation}

In an $n$-node system, the transition matrix is an $n \times n$ matrix. To construct a random matrix that satisfies detailed balance, we initialize the entries of the upper triangular part and the lower penultimate diagonal with values drawn from a Gaussian distribution with some mean (say, 1) and standard deviation (say, 0.3). Specifically, we sample independent transition rates as:
$$
k_{ij} = 1 + 0.3\  \eta_{ij}, 
$$
where $\eta_{ij}\sim\mathcal{N}(0,1)$ are random uncorrelated parameters drawn from a standard normal distribution.
The remaining entries in the lower triangular part are then determined by enforcing the detailed-balance constraint. For notational simplicity, we refer to $\tilde{k}_{ij}$ (as defined in the main text) simply as $k_{ij}$ in the appendices.

  \item \textbf{Breaking the detailed-balance condition:} To introduce non-equilibrium driving forces, we multiply one or more transition rates by exponential factors $e^{\mu}$, which represent light-transducing phototrophs--like batteries--in the ecology-inspired living circuits.  
    \item \textbf{Fast readjustment of electron densities (separation of timescales)}: The electron densities were assumed to equilibrate rapidly compared to the slower dynamics of adaptive variables $\Lambda_{ij}$, enabling a quasi-steady-state approximation. The dynamics of electron densities satisfy the Master equation, 
    \begin{equation}
\frac{dp_i}{dt} = \sum_j W_{ij} p_j \, ,
\end{equation}
where $W_{ij} = \Lambda_{ij} k_{ij}$ denotes the transition rate from state $j \rightarrow i $. Here, the $\Lambda_{ij} (= \Lambda_{ji})$ corresponds to the adaptive conductance, and $k_{ij}'s$ denote the non-adaptive elements chosen as discussed. The quasi-steady-state approximation implies that 
$$
\sum_{j}W_{ij}p_{j} = 0 \, . 
$$
This set of n equations forms a degenerate set of algebraic equations and hence, doesn't provide a unique solution. To obtain the solution, we take $(n-1)$ equations from the above set and supplement it with the conservation of probability condition, $\sum_{i}p_i = 1$. Solving the resultant set of equations gives the electron densities at network nodes as a function of $\Lambda_{ij}'$s. Since there are (n-1) equations that are collectively homogeneous of degree 1 in the variables $\Lambda'$s, we obtain all the $p'$s as functions of (at most) $(^{n}C_{2}-1)$ independent ratios of $\Lambda'$s.
    \item \textbf{Evolution of adaptive components in the living circuits}: The slow evolution of adaptive components, $\Lambda_{ij}$, is given by 
    \begin{equation}
\ \frac{d\Lambda_{ij}}{dt} = \sigma_{ij}(\vec{p}, \textbf{W}) - \sigma_{\text{maint}}\Lambda_{ij}
\end{equation}
where
\begin{equation}
    \sigma_{ij}(\vec{p}, \textbf{W}) = ( W_{ij} p_j - W_{ji} p_i) \log\left(\frac{W_{ij}p_j}{W_{ji}p_i}\right) \, .
\end{equation}
We note that the per-capita growth rate depends only on $p_{i}'$s i.e., the quasi steady-state electron densities. As discussed above, the fast dynamics of $p_i'$s let us determine them completely in terms of at most $(^{n}C_{2}-1)$ independent ratios of $\Lambda'$s. As a result, the steady state may not be reached, i.e., the $\Lambda'$s continue to increase without affecting the values of their ratios and the values of per capita edge dissipation. In other words, the dynamics remains invariant under scaling of the total conductance across all surviving edges. Therefore, to stop the simulations, we multiply the $\Lambda$ equations of the surviving  edges with a logistic term, i.e.,   
  \begin{equation}
\ \frac{d\Lambda_{ij}}{dt} = \Big(\sigma_{ij}(\vec{p}, \textbf{W}) - \sigma_{\text{maint}}\Lambda_{ij}\Big) \Big(1 - \alpha \sum_{i \neq j}\Lambda_{ij}\Big)
\end{equation}
where $\alpha$ is the carrying capacity. For simulations in this work, we take $\sigma_{\text{maint}} = 0.05$ and $\alpha = 10^{-4}$, resulting in a total conductance limit $\sum_{ij} \Lambda_{ij} = \frac{1}{\alpha} = 10^4$. Note that the specific choice of $\alpha$ does not affect the outcomes of our analysis. This is because the per-capita edge dissipation, as well as the normalized relative edge dissipations, converge to steady-state values irrespective of the inclusion of the logistic saturation term. Consequently, all reported results are inherently normalized with respect to the total carrying capacity, rendering the specific value of $\alpha$ inconsequential.
\end{enumerate}
\section{Mapping redox metabolic ecosystems to living circuits}
\label{mappingAppendix}

\begin{figure}[H]
    \centering
    \includegraphics[width=0.9\linewidth]{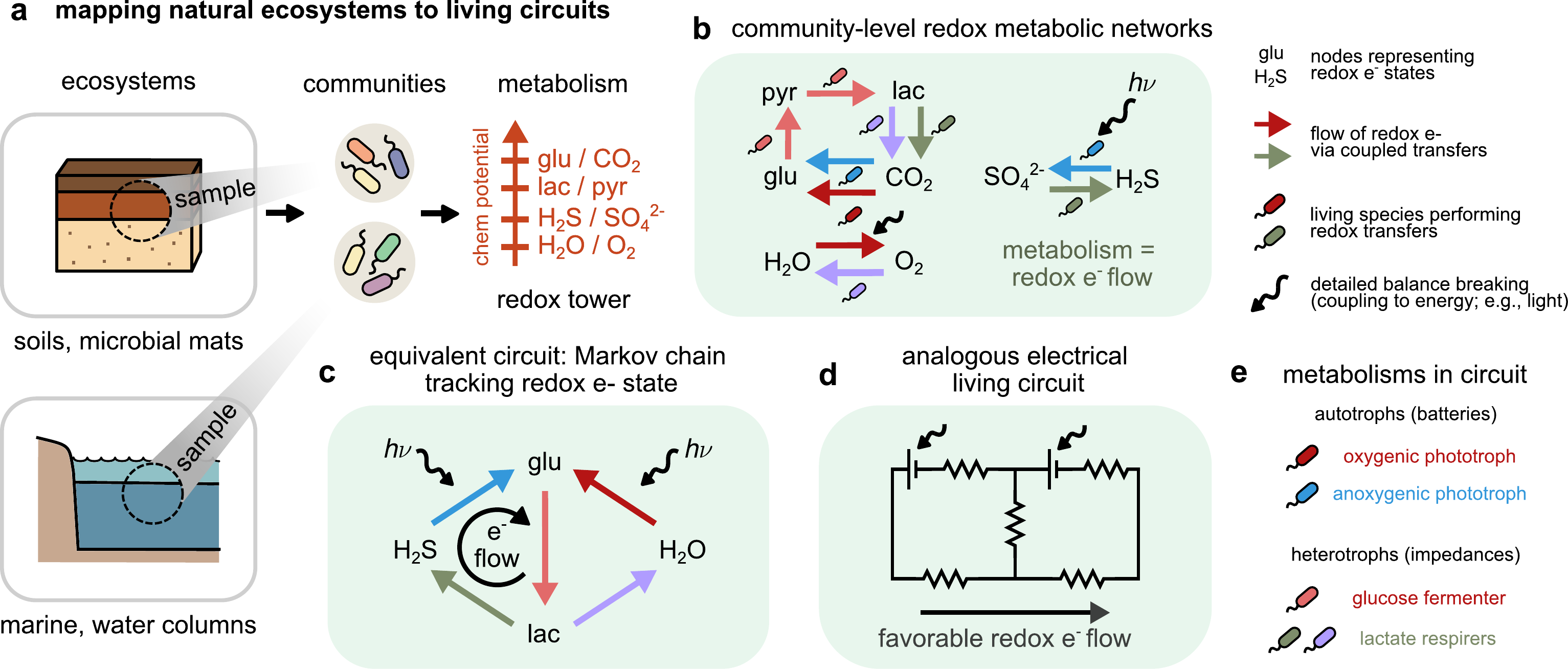}
\vskip 5 pt \caption{\textsf{\textbf{Mapping microbial ecosystems to adaptive living circuits.} (a) Each species catalyzes a net redox transformation, transferring electrons from \( R_i \to R_j \). (b) These transformations define directed edges in a Markov chain of redox states, representing metabolic activity across species. (c) The redox network maps onto an adaptive electrical circuit where species play the role of conductances that adjust based on local energy dissipation.}}
    \label{mapping_met_network_app_fig}
\end{figure}

We now detail our exact mapping from microbial redox ecosystems to adaptive electrical circuits (or equivalently, Markov chains), as illustrated in Fig.~\ref{mapping_met_network_app_fig}. While conceptually, the mapping from such ecosystems to living circuits (or Markov chains) is exact, not all the underlying parameters of the resulting circuits can be inferred directly from redox metabolism. As we show in the next section, redox metabolic potentials act as constraints on the parameters in the model while leaving some of them free. We detail choices on these free parameters in the next section (Appendix~\ref{redoxTowerAppendix}) 

Each microbial species performs a net redox transformation, coupling two half-reactions:
\[
R_i \to O_i + e^- \quad \text{and} \quad O_j + e^- \to R_j,
\]
thereby transferring electrons from state \( R_i \) to \( R_j \). In principle, species can perform two (or more) net reactions simultaneously, bifurcating electrons in multiple directions. In these cases, one may introduce more complex notation where those two or more edges represent the same underlying living species in the ecosystem. For simplicity we assume that each species only performs one net redox reaction.

\vskip 10pt
\textsf{\textbf{Redox states as Markov chain nodes.}} Coarse-graining the community-wide electron network in Fig.~\ref{mapping_met_network_app_fig}b where each species just connects the terminal redox states (reduced states in this case), we end up with a Markov chain representing interconversions between electrons in different redox states shown in Fig.~\ref{mapping_met_network_app_fig}c. For example, the pink species is a fermenter that uses glucose (glu) as a donor and lactose (lac) as a final acceptor, using pyruvate (pyr) as a in intermediate \cite{madigan1997brock,flamholz2025proteome}. Ultimately, this fermenter takes an electron on the reduced state glucose and converts it to one on the reduced state lactose. Each reduced redox state \( R_i \) is a node in a Markov chain, e.g., glucose (glu), lactose (lac) and H$_2$O as shown. A directed edge \( j \to i \) exists whenever one or more species catalyze the \textit{net} electron transfer from \( R_j \) to \( R_i \). The ecosystem-wide dynamics of electrons are thus described by a master equation:
\[
\frac{dp_i}{dt} = \sum_j W_{ij} p_j - \sum_j W_{ji} p_i,
\]
where \( W_{ij} \) is the total transition rate from \( j \to i \), aggregated across all species that catalyze this reaction.

\vskip 10pt
\textsf{\textbf{Ecosystem-wide metabolic interconversions as redox transitions.}} Each rate \( W_{ij} \) decomposes into:
\[
W_{ij} = \Lambda_{ij} k_{ij},
\]
where \( \Lambda_{ij} = \Lambda_{ji} \) is the abundance of species catalyzing the \( i \leftrightarrow j \) transformation, and \( k_{ij} \) is the spontaneous kinetic rate for the uncatalyzed reaction \( j \to i \). Species are interpreted as living catalysts that scale conductances, and thus the net reaction rates, based on their abundance.

\vskip 10pt
\textsf{\textbf{Ohm's law or linear approximation.}} We assume that metabolic reaction kinetics operate in the linear regime, i.e., as in Michaelis-Menten kinetics when substrate concentrations are much smaller than \( K_M \). This implies the net forward transformation rate is proportional to the substrate concentrations:
\[
J_{ij}^{\text{net}} \approx W_{ij} p_j - W_{ji} p_i = \Lambda_{ij} \left(k_{ij} p_j - k_{ji} p_i\right),
\]
leading to a linear Markov process. This is the analog of Ohm's law in our circuit analogy, where $J_{ij}^{\text{net}}$ is the net forward current, $\left(k_{ij} p_j - k_{ji} p_i\right)$ the net current due to chemical potential differences, with $\Lambda_{ij}$ being a multiplier to the conductance for the net reaction. Since $\Lambda_{ij}$ is analogous to species abundances, increasing would increase the net reaction rate since there would be more individuals of the species performing that reaction. Departures from this linear regime would yield non-linear dynamics analogous to circuits with non-Ohmic elements.

\vskip 10pt
\textsf{\textbf{Detailed balance breaking.}} To sustain non-equilibrium steady states (NESS), some reactions are powered by external energy sources, such as photons or chemical disequilibria. This is captured by:
\[
k_{ij} = \tilde{k}_{ij} e^{\phi_{ij}}, \quad \text{so that} \quad \frac{k_{ij}}{k_{ji}} = \left( \frac{\tilde{k}_{ij}}{\tilde{k}_{ji}} \right) e^{\mu},
\]
where \( \phi_{ij} \) represents the energy input (e.g., light in photosynthesis), and \( \mu = \phi_{ij} - \phi_{ji}\) quantifies the nonequilibrium drive. This breaks detailed balance and generates persistent electron currents.

\vskip 10pt
\textsf{\textbf{Multiple paths between nodes.}} Some pairs of nodes are connected by multiple metabolic strategies. For example, H\(_2\text{O} \leftrightarrow \text{glucose} \) could be catalyzed either by phototrophs (photosynthesis) or heterotrophs (aerobic respiration). These correspond to distinct paths. Standard Markov chains cannot encode such multiplicity without intermediate nodes \cite{schnakenberg1976network}. Including both metabolisms in the same circuit can be resolved this by introducing intermediate nodes (e.g., carriers such as NADH in different contexts), creating distinct redox states to separate pathways. For simplicity and to focus on phenomenology, in this manuscript we avoid this complication and instead only allow one path for each pair of nodes.

\section{Parameterizing living circuits using redox thermodynamics data}
\label{redoxTowerAppendix}

\begin{figure*}
    \centering
    \includegraphics[width=0.9\linewidth]{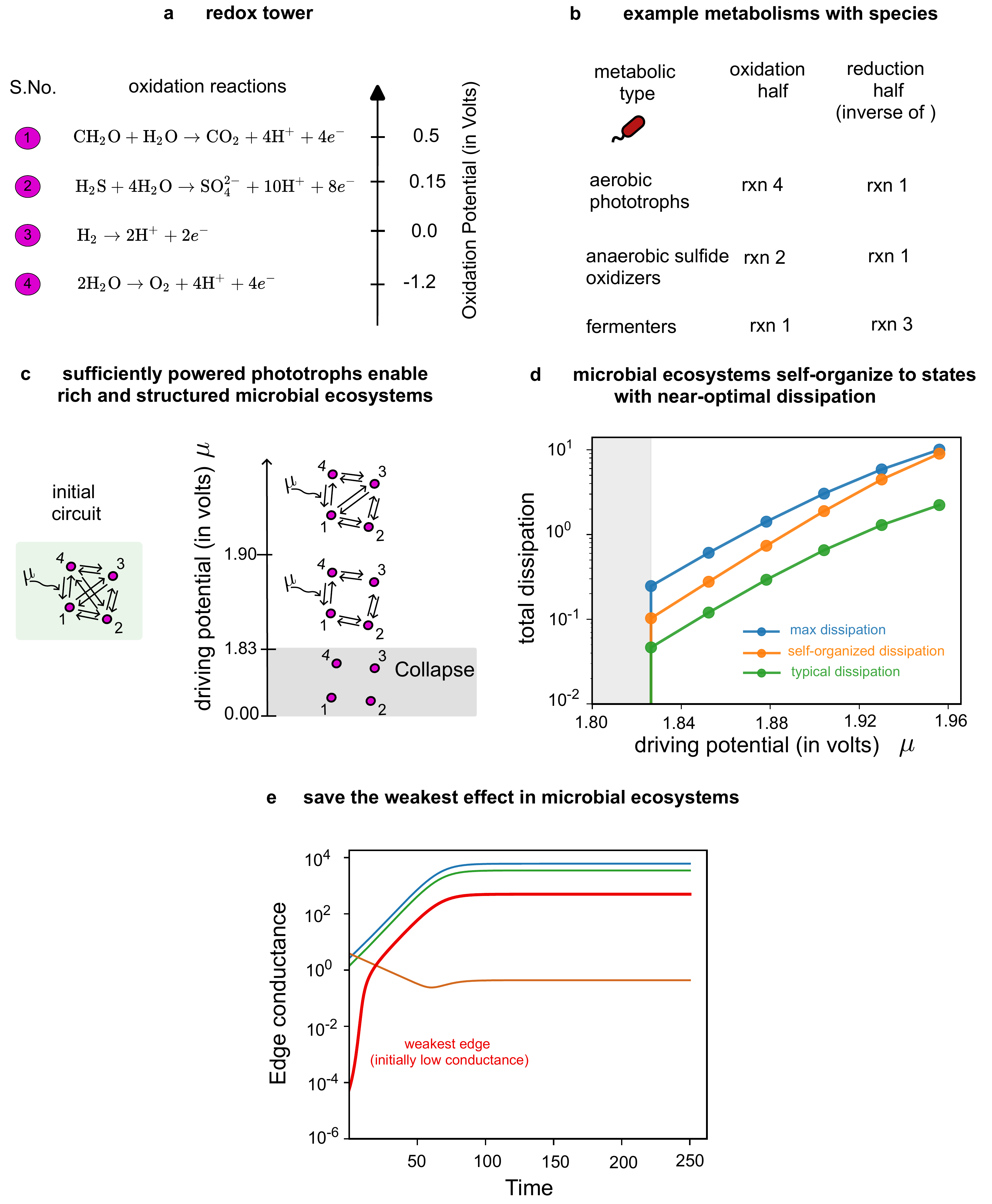}
\caption{\justifying\textsf{\textbf{Parametrizing living circuits using redox thermodynamics.} (a) A set of reactions from the redox tower with their chemical potentials (re-scaled relative to a chosen reference of reaction 3). In this convention a higher redox potential indicates a greater propensity to donate electrons. Values for oxidation potential are estimates from \cite{madigan1997brock}. (b) Examples of microbial species in a an ecosystem which couple distinct redox reactions and their corresponding metabolic and ecological roles. (c) Results of simulations of living circuit dynamics corresponding to the microbial community of species in (b), i.e., with $k_{ij}$'s parametrized using the potentials in (a). Analogous to Fig.~\ref{fig5}a, we show the final steady-state circuit topologies when simulated with increasing nonequilibrium drive $\mu$. Similar to random $k_{ij}$, we see similar results: (c) shows a critical drive $\mu_{\text{crit}} \sim 1.83$ V and increasing network complexity at greater values of the drive. (d) shows that for $\mu > \mu_{\text{crit}}$, circuits settle into steady-states with near-maximal total dissipation (circuit dissipation in orange compared with theoretical maximum in blue), much greater than typical (with randomized conductances shown in green); this is consistent with Fig.~\ref{fig4} in the main text with randomized $k_{ij}$'s. (e) Similarly, this circuit also exhibits a ``save the weakest'' effect, where the weakest edge (red) with initial conductance $10^4$ times lower than the rest can be rescued from extinction, akin to Fig.~\ref{fig3} with random $k_{ij}$.}}
    \label{real_redox_tower}
\end{figure*}
Consider a set of naturally occurring redox half-reactions, each characterized by a redox potential $\mu_i$ (Fig.~\ref{real_redox_tower}a). In biological species, specialized enzymes facilitate electron transfer in redox reactions, thereby modulating the kinetics of the redox metabolic network. Without such enzymes, the redox potentials $\mu_i$ determine a thermodynamic landscape that governs the equilibrium distribution of electrons among redox states.

At equilibrium, the distribution of electrons among different redox pairs satisfies detailed balance, with the ratio of forward to backward rates given by \cite{qian2007phosphorylation}:

\begin{equation}
\frac{k_{ij}}{k_{ji}} = e^{\beta (\mu_j - \mu_i)} \, ,
\end{equation}
where $\beta = \frac{1}{k_B T}$ is the inverse thermal energy.

While the redox potentials constrain the \emph{ratios} of the transition rates, they do not uniquely determine the \emph{absolute values} of the rates. To resolve this, we have to include activation energies due to kinetic barriers between states. To estimate and include these into our kinetics, we have to make some choices. One reasonable choice is to set some activation energies proportional to the energy difference between states, while respecting the relative ease of forward and reverse transitions as well as detailed balance. Another choice is to sample them from some distribution. In an example shown here, we make the former choice.

Specifically, we prescribe a consistent assignment of barrier heights based on the following rule. For a redox transition from state $i$ to state $j$, if $\mu_i < \mu_j$, we define:
\begin{equation}
k_{ij} = k_0 \, e^{- \frac{1}{2} \beta (\mu_j - \mu_i)} \, , \hspace{2cm}
k_{ji} = k_0 \, e^{- \frac{3}{2} \beta (\mu_j - \mu_i)} \, .
\label{redox DB matrix}
\end{equation}
This prescription ensures that the barrier for the \emph{forward transition} lies at a height $\frac{\mu_{j}-\mu_{i}}{2}$ from the higher potential state, while the \emph{reverse transition} has a correspondingly higher barrier at a height $\frac{3(\mu_{j}-\mu_{i})}{2}$ from the lower potential state, thereby satisfying detailed balance.

In the living circuit counterpart of these reactions, microbial species grow by extracting energy through the coupling two such redox reactions, thereby defining their functional roles or metabolic type within the ecosystem (Fig.~\ref{real_redox_tower}b). To simulate these living circuits, we take the detailed-balance transition matrix of $k_{ij}$'s as given in \ref{redox DB matrix}.
Once we have this non-adaptive part of the transition matrix, we follow steps $(2)-(4 )$ in Appendix~\ref{Methods} to parametrize the living circuits with adaptive dynamics (Fig.~ \ref{real_redox_tower}). We take $k_{0} \approx 10^{10}\ \text{s}^{-1}$, $\beta = 1/2.5$ Volt$^{-1}$ for room temperature and $\sigma_{\text{maint}} = 5 \times 10^{-2}$. These parameters are chosen arbitrarily.

Similar to results with randomly chosen $k_{ij}$ as shown in the main text (see Appendix~\ref{Methods} for Methods), we find that the circuits parameterized using chemical potentials from a redox tower (Fig.~\ref{real_redox_tower}) show the same qualitative results as the core phenomena we report: (1) the existence of a critical driving potential $\mu_{\text{crit}}$ above which circuits reach a dissipative NESS; (2) near-optimal dissipation, (3) an increasing drive-complexity relationship and (4) a ``save the weakest'' effect. In this ecosystem analog, the edge that couples to the nonequilibrium drive $\mu$ is a phototroph, which transfers electrons from $\text{H}_2\text{O}$ to $\text{CH}_2\text{O}$.

\section{Critical driving potential for NESS}

Here we derive an analytical expression for the critical driving potential $\mu_{\text{crit}}$ required for a living circuit to sustain a non-equilibrium steady state (NESS). This threshold marks a first-order phase transition between network collapse and functional self-organization.

\label{critmuappendix}

First, consider a 3-cycle where the per capita dissipation along any edge—for instance, the edge connecting nodes 
$1$ and $2$—must be strictly positive for that edge to persist, i.e., 
\begin{equation}
\frac{\Lambda_{13}\Lambda_{23}}{B}(e^{\mu} - 1)k_{21}k_{13}k_{32} \log A - \sigma_{\text{maint}} > 0,
\end{equation}
where 
$$ A = \frac{k_{12}e^{\mu}(\Lambda_{23}\Lambda_{13} k_{23}W_{31} + \Lambda_{13}\Lambda_{12} k_{13}k_{21} + \Lambda_{23}\Lambda_{12} k_{23}k_{21})}{k_{21}(\Lambda_{13}\Lambda_{12} k_{13}k_{12} + \Lambda_{13}\Lambda_{23} k_{13}k_{32} + \Lambda_{12}\Lambda_{23} k_{12}k_{23})} $$
and 
\begin{eqnarray} B &=&\Lambda_{13}\Lambda_{12} k_{13}k_{12}e^{\mu} + \Lambda_{13}\Lambda_{23} k_{13}k_{32} + \Lambda_{12}\Lambda_{23} k_{12}k_{23}e^{\mu}  \nonumber \\ & & + \Lambda_{23}\Lambda_{13} k_{23}k_{31} + \Lambda_{13}\Lambda_{12} k_{13}k_{21} + \Lambda_{23}\Lambda_{12} k_{23}k_{21} \nonumber \\ & & + \Lambda_{23}\Lambda_{13} k_{32}k_{31}  + \Lambda_{23}\Lambda_{12} k_{32}k_{21} + \Lambda_{12}\Lambda_{13} k_{31}k_{12}e^{\mu}  \, .\nonumber
\end{eqnarray}

Since all terms in the above expression are homogeneous of the same order in the $\Lambda'$s in both the numerator and the denominator, the dynamics on any simplex of edge conductances can be rescaled and mapped onto the unit simplex (see Appendix \ref{Methods}). In particular, approaching the extinction of an edge - e.g., when $\Lambda_{12} = \varepsilon \to 0$ - implies that the conductances of the remaining two edges remain finite and non-zero. Therefore, near extinction, the edge dissipation is 
\begin{equation}
\frac{1}{B_{0}}(e^{\mu} - 1)k_{21}k_{13}k_{32} \log A_{0} - \sigma_{\text{maint}} > 0
\end{equation}
where $B_0$ and $A_{0}$ is independent of $\Lambda's$:
\begin{align}
B_0 &= (k_{13}k_{32} + k_{23}k_{31} +  k_{32}k_{31} )
\end{align}
and 
$$
 A_{0} = e^{\mu}\frac{k_{12} k_{23}k_{31}}{k_{21}k_{13}k_{32}} = e^{\mu} .
$$




As $\mu$ increases, the first term grows, eventually balancing $\sigma_{\text{maint}}$ at the critical point:

\begin{equation}
\frac{1}{B_0}(e^{\mu_{\text{crit}}} - 1)k_{21}k_{13}k_{32} \log A_{0} = \sigma_{\text{maint}}
\end{equation}

This implies:

\begin{equation}
\big(e^{\mu_{\text{crit}}}-1\big)\mu_{\text{crit}} \approx \frac{\sigma_{\text{maint}} B_0}{k_{21}k_{13}k_{32}}
\end{equation}
 This critical value is independent of $\varepsilon$, confirming that the threshold for maintaining a NESS depends only on the network's kinetics and topology, not on the absolute magnitude of the conductances when they are all scaled equally. 


We can also extend this analysis to $N$-node cycles, where nodes are labeled $0, 1, 2, \ldots, N-1$ and edge $(N-1)-0$ couples to the external energy source. Following similar arguments, the critical driving potential for an N-node cycle is:

\begin{equation}
\big(e^{\mu_{\text{crit}}}-1\big)\mu_{\text{crit}} \approx \frac{\sigma_{\text{maint}}}{\chi_N(k)}
\end{equation}

where $\chi_N(k)$ is the emergent current for the N-node case:

\begin{equation}
\chi_N(k) = \frac{\prod_{i=0}^{N-1} k_{i,i+1}}{B_N}
\end{equation}

Here, $B_N$ is a generalized form of $B_0$ for N nodes:

\begin{equation}
B_N = \sum_{i,j} k_{ij}k_{ji} \cdot C_{ij}(e^{\mu})
\end{equation}

where $C_{ij}(e^{\mu})$ is either 1 or $e^{\mu}$ depending on whether the path from $i$ to $j$ includes the energy-coupled edge.



This general expression reveals a common feature: the critical driving potential $\mu_{\text{crit}}$ is inversely proportional to an emergent current $\chi_N$ that depends on both the topology and kinetics of the circuit, but is independent of the absolute magnitude of the edge conductances when they are all scaled equally.

\section{First-order phase transition at non-zero threshold in adaptive living circuits}
\label{1storderappendix}
The dynamics of living circuits, much like those of conventional electrical circuits, can be described as a random walk of electrons across the nodes of a bidirectional network:
\begin{equation}
\frac{dp_i}{dt} = \sum_j W_{ij} p_j \, ,
\end{equation}
where $W_{ij} $ denotes the transition rate from state $j \rightarrow i $. Unlike classical Markov processes, however, living circuits are characterized by adaptive transition rates—meaning $ W_{ij} $ is not fixed but evolves in time in response to the instantaneous probability distribution $ \vec{p} $ over the network:
\begin{equation}
\frac{dW_{ij}}{dt} = f(\vec{p}) \, .
\end{equation}
The function $f(\vec{p})$, which governs the adaptive dynamics, is context-dependent and can vary significantly depending on the underlying biological or physical scenario. One illustrative example is Hebbian learning, where synaptic connections between neurons strengthen in response to simultaneous activity, effectively allowing the network to adapt based on past stimuli.

In ecosystem-inspired circuit models, the transition rate decomposes as
$$
W_{ij} = \Lambda_{ij} k_{ij} \, ,
$$
where $k_{ij}$ is the baseline (static) transition rate between nodes $i$ and $j$, and $\Lambda_{ij}$ represents an adaptive factor—specifically, the abundance of a species that facilitates electron transitions across the $i$-$j$ edge. Since these species enhance conductance, $\Lambda_{ij}$ can be interpreted as the analog of edge conductance in electrical circuits. As these species harness the energy released during electron transitions to both sustain the structural machinery required for facilitation of electron transition and support their growth, the adaptive component or the edge conductance evolves as   

\begin{equation}
\ \frac{d\Lambda_{ij}}{dt} = \sigma_{ij}(\vec{p}, \textbf{W}) - \sigma_{\text{maint}}\Lambda_{ij}
\end{equation}
where:
\begin{equation}
    \sigma_{ij}(\vec{p}, \textbf{W}) = ( W_{ij} p_j - W_{ji} p_i) \log\left(\frac{W_{ij}p_j}{W_{ji}p_i}\right)
\end{equation} represents energy dissipation along the $ij$ edge, and $\sigma_{\text{maint}}$ is the minimum dissipation required to maintain the energy-harvesting structure. The need for a structural machinery and its associated cost imply:
\begin{itemize}
    \item The system requires a threshold amount of energy input to sustain non-equilibrium steady states. 
    \item  Any NESS must dissipate energy at least equal to the cost of maintaining the structural machinery i.e., NESS dissipating arbitrarily small energy cannot exist.
\end{itemize} To investigate these predictions, we consider the simplest possible nontrivial topology: a 3-cycle—three nodes connected in a loop with one species per edge facilitating transitions.

Before specializing to this case, we note that for an n-cycle, the non-equilibrium steady state distribution is given by 
\begin{equation}
p_{k} = \frac{T_{k} S_{n+1} + (1-T_{n+1})S_{k}}{\mathcal{Z}}
\end{equation}
where $$ \mathcal{Z} = \sum_{k=1}^{n}\Big(T_{k} S_{n+1} + (1-T_{n+1})S_{k} \Big)$$ 
and 
$$T_{k} = \prod_{i=1}^{k-1} \frac{W_{i+1,i}}{W_{i,i+1}} \  \ \ \text{k=2,3,..,n+1}\ \ \ \text{and}\ \ \  T_{1}=1 \, ,$$
$$S_{k} = \sum_{m=1}^{k-1}\prod_{i=m+1}^{k-1}\frac{W_{i+1,i}}{W_{i,i+1}}\frac{1}{W_{m,m+1}}
\  \ \ \text{k=2,3,..,n+1} \ \ \ \text{and} \ \ \ S_{1}=0 \, .$$
The total dissipation through edge $i\to (i+1)$ is given by   
\begin{equation}
\sigma_{i,i+1} =(W_{i,i+1}p_{i+1}-W_{i+1,i}p_{i})\log \Big(\frac{W_{i,i+1}p_{i+1}}{W_{i+1,i}p_{i}}\Big) \, .
\end{equation}
For living circuits with dynamical conductances, i.e., substituting $W_{i,i+1}\to \Lambda_{i,i+1}k_{i,i+1}$, this becomes
\begin{eqnarray}
\sigma_{i,i+1}
&=&  \Lambda_{i,i+1}(k_{i,i+1}p_{i+1}-k_{i+1,i}p_{i})\log \Big(\frac{k_{i,i+1}p_{i+1}}{k_{i+1,i}p_{i}}\Big) \, .\hspace{0.5cm}
\end{eqnarray}
Using the steady-state distribution, 
\begin{flushleft}
\begin{align}
\hspace{-0.5cm}\mathcal{\sigma}_{i,i+1} &=& \Lambda_{i,i+1}\Bigg(\frac{(k_{i,i+1}T_{i+1}-k_{i+1,i}T_{i}) S_{n+1}}{\mathcal{Z}} +   \frac{(1-T_{n+1})(k_{i,i+1}S_{i+1}-k_{i+1,i}S_{i})}{\mathcal{Z}}\Bigg)\log \Bigg(\frac{k_{i,i+1}(T_{i+1} S_{n+1} + (1-T_{n+1})S_{i+1})}{k_{i+1,i}(T_{i} S_{n+1} + (1-T_{n+1})S_{i})}\Bigg) \nonumber \\
&= & \frac{(1-T_{n+1})}{\mathcal{Z}}\log \Bigg(\frac{k_{i,i+1}p_{i+1}}{k_{i+1,i}p_{i}}\Bigg) \hspace{12cm}
\end{align}
\end{flushleft}
This implies that 

$$
\sigma_{\text{tot}} = \sum_{i}\sigma_{i,i+1} = \frac{(1-T_{n+1})}{\mathcal{Z}} \log T_{n+1}
$$
and therefore, a non-equilibrium steady state requires breaking the detailed balance condition, 
$$
\log\Bigg(\frac{W_{21}W_{32}... W_{1n}}{W_{12} W_{23}...W_{n1}}\Bigg) = \log\Bigg(\frac{k_{21}k_{32}... k_{1n}}{k_{12} k_{23}...k_{n1}}\Bigg) =  e^{\mu}\neq 1  \, .
$$
We implement this in the ecosystem-inspired circuit by coupling one edge (say 0 - 1 ) to an external source of energy such as light: $k_{01} \to k_{01}e^{\mu} $.

\begin{figure}[h]
\centering\includegraphics[width=0.7\linewidth]{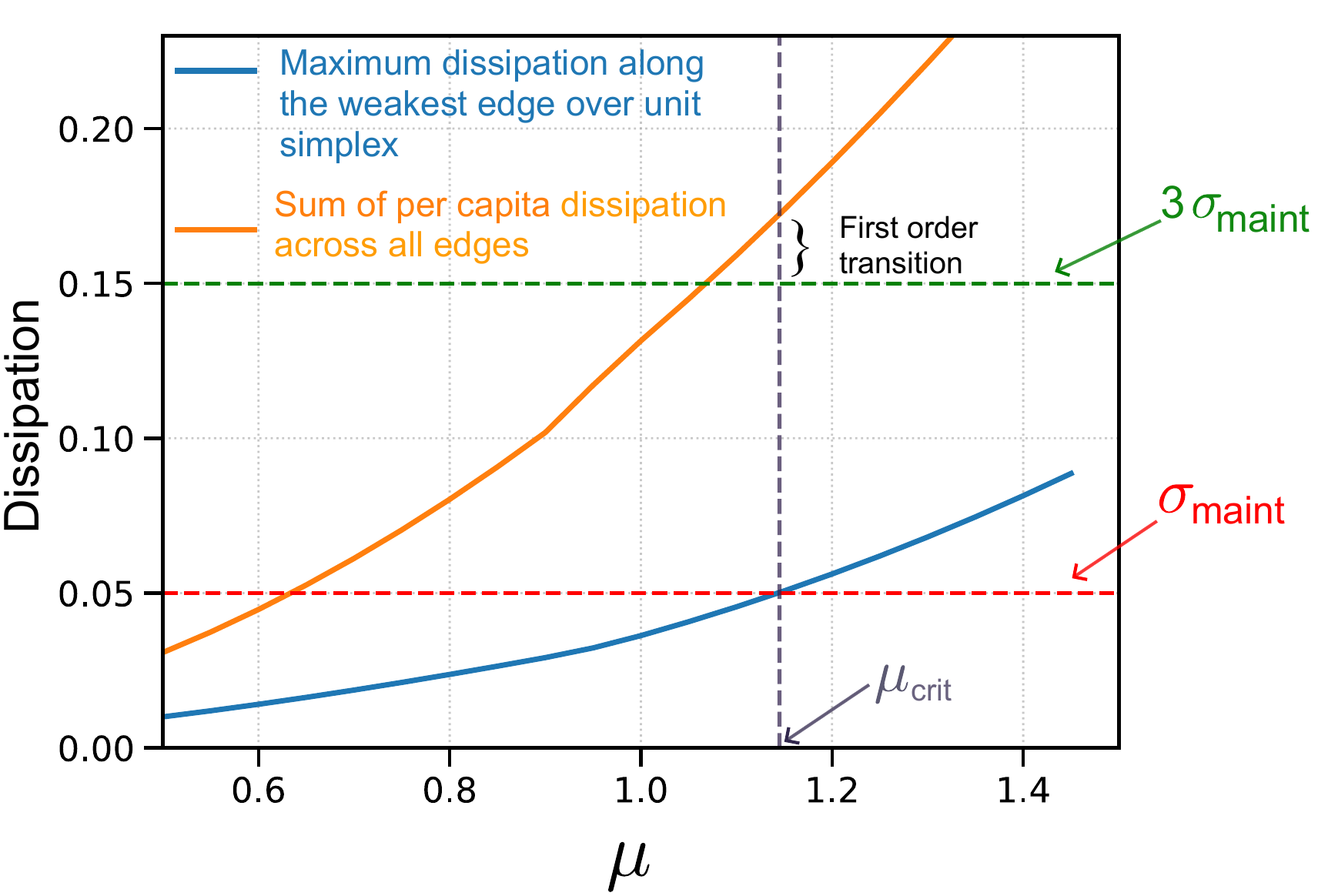} \vspace{0.1cm}
\caption{\justifying \textsf{\textbf{Critical energy input and first-order phase transition in a 3-node adaptive living circuit.} The blue curve shows the maximum possible dissipation along the weakest edge across all configurations on the conductance simplex. Since the species facilitating transitions require a minimum energy $\sigma_{\text{maint}}$ to sustain their energy-harvesting machinery, the circuit can maintain a non-equilibrium steady state (NESS) only when the blue curve lies above the $\sigma_{\text{maint}}$ threshold. The red curve represents the total per capita dissipation summed over all three edges. At the critical energy input $\mu_{h\nu}^{\text{critical}}$, the red curve lies significantly above the $3\sigma_{\text{maint}}$ line, indicating that the NESS dissipates more energy than the bare minimum required for survival. These curves correspond to a specific realization of the baseline rates $k_{ij}$, and while the precise values may vary across realizations, the qualitative behavior remains unchanged.}}
    \label{fig:enter-label}
\end{figure}

In a 3-node network, a sustainable non-equilibrium steady state (NESS) requires that all three edges dissipate more than $\sigma_{\text{maint}}$, otherwise the corresponding conductance $\Lambda_{ij}$ will decay to zero. To identify the critical energy input $\mu_{\text{crit}}$, we search the conductance simplex for configurations where the minimum edge dissipation exceeds $\sigma_{\text{maint}}$. Such configurations exist only if $\mu > \mu_{\text{crit}} > 0$, below which all living circuits collapse.

Moreover, at $\mu_{\text{crit}}$, we confirm that the total dissipation exceeds the minimum required for survival:
\begin{equation}
\sum_{\text{edges}} \frac{\sigma_{ij}}{\Lambda_{ij}}> 3 \sigma_{\text{maint}} \, ,
\end{equation}
indicating a first-order phase transition. The value of $\mu_{\text{crit}}$, as well as the residual dissipation above the minimum required, are determined by the baseline rates $k_{ij}$.
\subsection{Limit $\Lambda_{i,i+1} \to 0$}
We now consider a network configuration in which one of the edges is about to go extinct, i.e., $\Lambda_{i,i+1} \approx 0$, and is much smaller compared to the conductances of all other edges. In this limit, 
\begin{equation}
    S_{l (\leq i)} << S_{k (\geq i+1)} \approx \frac{T_{k}}{T_{i+1}} \frac{1}{k_{i,i+1} \Lambda_{i,i+1}}
\end{equation}
Similarly, we also find that 
\begin{equation}
\mathcal{Z} \approx \frac{1}{k_{i,i+1}\Lambda_{i,i+1}} \frac{1}{T_{i+1}} \Big[ T_{n+1} \sum_{l=1}^{k}T_{l} + \sum_{l=k+1}^{n}T_{l}\Big]
\end{equation}
Using these expressions, the dissipation through the dying edge is given by 
\begin{eqnarray}
    \sigma_{i,i+1} &=& \frac{(1-T_{n+1})}{ \frac{1}{k_{i,i+1}\Lambda_{i,i+1}} \frac{1}{T_{i+1}} \Big[ T_{n+1} \sum_{l=1}^{k}T_{l} + \sum_{l=k+1}^{n}T_{l}\Big] + O(1)} \log\Bigg(\frac{1/\Lambda_{i,i+1} + O(1)}{T_{n+1}/\Lambda_{i,i+1} + O(1)}\Bigg) \nonumber \\
    &=& \frac{(1-T_{n+1})T_{i+1} k_{i,i+1}\Lambda_{i,i+1}}{  \Big[ T_{n+1} \sum_{l=1}^{k}T_{l} + \sum_{l=k+1}^{n}T_{l}\Big] + O(\Lambda_{i,i+1})} \log\Bigg(\frac{1 + O(\Lambda_{i,i+1})}{T_{n+1} + O(\Lambda_{i,i+1})}\Bigg) \nonumber \\
    &=& \frac{(T_{n+1}-1)T_{i+1} k_{i,i+1}\Lambda_{i,i+1}}{  \Big[ T_{n+1} \sum_{l=1}^{k}T_{l} + \sum_{l=k+1}^{n}T_{l}\Big]} \Big(\log(T_{n+1}) +  O(\Lambda_{i,i+1})  
     + \log(1 + O(\Lambda_{i,i+1}))\Big)
\end{eqnarray}
whereas dissipation through other edges is given by 
\begin{equation}
    \sigma_{k,k+1} = 0 + \Lambda_{i,i+1} O(\Lambda_{i,i+1}) \, .
\end{equation}

\section{Save the weakest: approximate Lyapunov functional near edge extinction}
\label{xlogxappendix}
From the above analysis, we find that the dynamics of edge conductances during approach towards the edges of the conductance simplex, to leading order, is given by  
\begin{equation}
 \dot{\Lambda}_{ij} = -\frac{\delta \mathcal{F}}{\delta \Lambda_{ij}}   
\end{equation}
where 
\begin{equation}
    \mathcal{F} = -\sum_{i=1}^{n-1} \Bigg(\frac{(T_{n+1}-1)T_{i+1} k_{i,i+1}}{ T_{n+1} \sum_{l=1}^{k}T_{l} + \sum_{l=k+1}^{n}T_{l}} \log(T_{n+1})-\sigma_{\text{maint}}\Bigg) \frac{\Lambda_{i,i+1}^{2}}{2}
\end{equation}

\subsection{Limit $\Lambda_{i,i+1} \to \infty$} 
It is easy to see that, in the limit of infinite conductance through the edge $i-(i+1)$, the dissipation through this edge is 0, effectively severing the 
n-cycle. Thus, the Lyapunov functional during approach towards infinitely large conductances is given by 
\begin{equation}
\mathcal{F} = \sum_{i=1}^{n-1} \sigma_{\text{maint}}\frac{\Lambda_{i,i+1}^{2}}{2}    
\end{equation}

\section{Time varying environments}
\label{timeVarEnvAppendix}
\begin{figure}[ht!]
\centering
\includegraphics[width=0.7\linewidth]{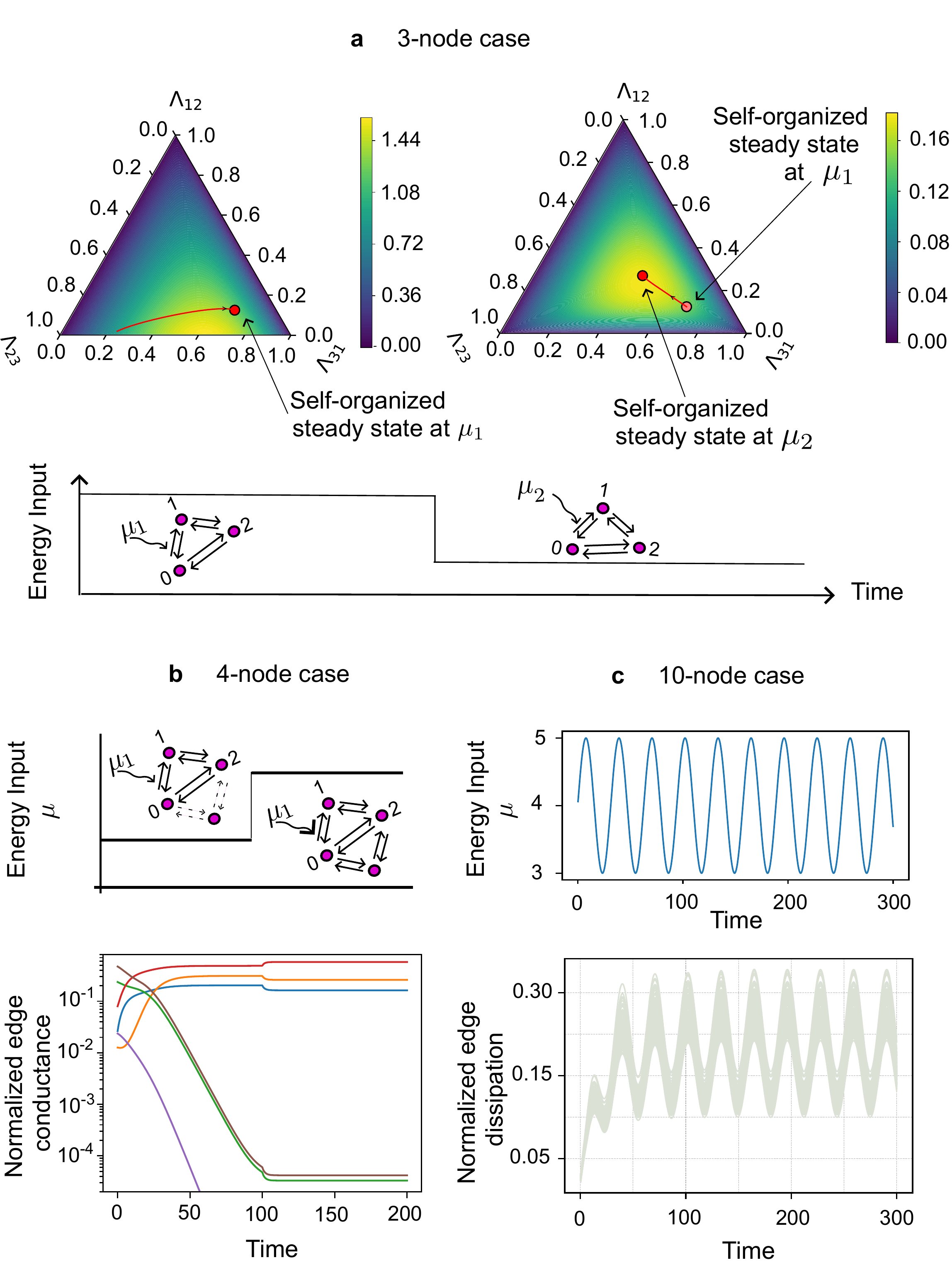}
\vspace{0.1cm}
    \caption{\justifying \textsf{\textbf{Adaptability of living circuits to time varying environments.}  Changing the external energy input of the living circuits forces them to adjust the electron densities over the Markovian states, affecting both the ability of circuits to dissipate as well as their topological structure. (a) When the external energy input to a 3-node cycle is suddenly decreased, the circuit adjusts electron densities across nodes to reduce total dissipation. (b) Conversely, increasing the energy input in a 4-node network prevents the imminent extinction of two edges, allowing the circuit to stabilize into a NESS with a more intricate topology—effectively healing the topological scar induced by the previously low-energy environment. (c) In a 10-node network, the circuit continuously adapts its local dissipation profile in response to a sinusoidally varying external energy input, demonstrating dynamic tuning of its internal structure.}}
    \label{timevarfigre}
\end{figure}
\noindent
In the main text, we show that for a given set of bare $k_{ij}\ '$s and external drive, living circuits reliably converge to the same final state, regardless of their initial conditions. This demonstrates the robustness of living circuits under constant environmental conditions. Remarkably, we also find that even when the environment changes, living circuits adapt seamlessly, reaching the new steady state without retaining any memory of the previous conditions. This behavior can be intuitively understood: the steady state attained under the earlier environment simply becomes the initial condition for the new one, and given sufficient time, the circuit fully adjusts to the new conditions, effectively erasing any trace of its prior state (see Fig. \ref{timevarfigre}). 

\section{Deviation from global maximization of total dissipation by living circuit dynamics}
\label{gradDesAppendix}
In the above section, we see that the total dissipation through an n-cycle is given by : 
$$
\sigma_{\text{tot}} = \sum_{i}\sigma_{i,i+1} = \frac{(1-T_{n+1})}{\mathcal{Z}} \log T_{n+1} \, .
$$
The gradient ascent dynamics should thus be given by 
$$ 
\dot{\Lambda}_{i,(i+1)}^{\text{GA}} = \frac{\delta \sigma_{\text{tot}}}{\delta \Lambda_{i,(i+1)}}
 = -\sigma_{\text{tot}}\frac{\delta \log \mathcal{Z}}{\delta \Lambda_{i,(i+1)}}
$$
whereas the actual dynamics is 
\begin{eqnarray}
    \dot{\Lambda}^{\text{Actual}}_{i,(i+1)}  &=& \sigma_{i,(i+1)} - \sigma_{maint} \Lambda_{i,(i+1)}  =  \frac{(1-T_{n+1})}{\mathcal{Z}}\log \Bigg(\frac{k_{i,i+1}p_{i+1}}{k_{i+1,i}p_{i}}\Bigg) - \sigma_{\text{maint}} \Lambda_{i,(i+1)}  \nonumber \\
    & = & \frac{(1-T_{n+1})}{\mathcal{Z}} \log \Bigg(1 + \frac{(1-T_{n+1})}{\Lambda_{i,(i+1)} k_{i+1,i}(T_{i} S_{n+1} + (1-T_{n+1})S_{i})}\Bigg)  - \sigma_{\text{maint}} \Lambda_{i,(i+1)} \nonumber \\ 
    & \approx & \frac{(1-T_{n+1})^{2}}{\Lambda_{i,(i+1)} k_{i+1,i}(T_{i} S_{n+1} + (1-T_{n+1})S_{i})}\frac{\frac{\delta\log \mathcal{Z}}{\delta \Lambda_{i,(i+1)}}}{\frac{\delta \mathcal{Z}}{\delta \Lambda_{i,(i+1)}}}- \sigma_{\text{maint}} \Lambda_{i,(i+1)} 
\end{eqnarray}
Since $\mathcal{Z} = \sum_{i=1}^{n-1} \frac{C_{i}}{\Lambda_{i,i+1}}$
$$ 
\frac{\dot{\Lambda}^{\text{Actual}}_{i,(i+1)}}{\Lambda_{i,(i+1)}}  = \frac{(1-T_{n+1})^{2}}{ C_{i}k_{i+1,i}(T_{i} S_{n+1} + (1-T_{n+1})S_{i})}\frac{1}{\sigma_{\text{tot}}}\underbrace{\frac{\delta \sigma_{\text{tot}}}{\delta \Lambda_{i,(i+1)}}}_{\text{Gradient ascent}}- \sigma_{\text{maint}}
$$
Therefore, the actual dynamics of living circuits differ from the gradient ascent dynamics through the $\sigma_{\text{maint}}$ term as well as through the coefficient of the $\sigma_{\text{tot}}-$derivative in the first term. 


\section{Adaptive resistance dynamics from minimum dissipation}
\label{passiveMinDissDynamics}
We derive a local adaptive rule for edge resistances \( R_e \) in an electrical resistor network where currents \( I_e \) obey Kirchhoff's current law (KCL) and adjust instantaneously to minimize the total power dissipation. We assume that external sources and sinks provide a fixed divergence condition \( \nabla \cdot \mathbf{I} = \mathbf{s} \), with \( \mathbf{s} \) independent of time. Internal currents relax rapidly compared to resistance changes and are always at their constrained dissipation-minimizing configuration. The system seeks to adapt resistances \( \{ R_e \} \) to minimize total dissipation, subject to a cost constraint.

Let the network be represented as a directed graph with edge set \( \mathcal{E} \), node set \( \mathcal{V} \), and incidence matrix \( A \in \mathbb{R}^{|\mathcal{V}| \times |\mathcal{E}|} \). Define \( \mathbf{I} \in \mathbb{R}^{|\mathcal{E}|} \) as the vector of edge currents and \( \mathbf{R} \in \mathbb{R}^{|\mathcal{E}|} \) as the vector of edge resistances. The total dissipated power is

\begin{equation}
P[\mathbf{I}, \mathbf{R}] = \sum_{e \in \mathcal{E}} R_e I_e^2.
\end{equation}

For fixed resistances, the currents minimize dissipation subject to KCL:

\begin{equation}
\min_{\mathbf{I}} P[\mathbf{I}, \mathbf{R}] \quad \text{subject to} \quad A \mathbf{I} = \mathbf{s}.
\end{equation}

To enforce the constraint, we introduce Lagrange multipliers \( \boldsymbol{\lambda} \in \mathbb{R}^{|\mathcal{V}|} \) and define the Lagrangian

\begin{equation}
\mathcal{L}(\mathbf{I}, \boldsymbol{\lambda}; \mathbf{R}) = \sum_e R_e I_e^2 + \boldsymbol{\lambda}^\top (A \mathbf{I} - \mathbf{s}).
\end{equation}

Stationarity requires

\begin{align}
\frac{\partial \mathcal{L}}{\partial I_e} &= 2 R_e I_e + \sum_v \lambda_v A_{ve} = 0, \label{eq:kkt-current} \\
\frac{\partial \mathcal{L}}{\partial \lambda_v} &= (A \mathbf{I} - \mathbf{s})_v = 0. \label{eq:kkt-kcl}
\end{align}

Equations \eqref{eq:kkt-current} and \eqref{eq:kkt-kcl} define the constrained minimizing current configuration \( \mathbf{I}[\mathbf{R}] \).

We now compute the total derivative of dissipation with respect to resistance. Since \( \mathbf{I} \) depends on \( \mathbf{R} \) through the constrained minimization, we compute

\begin{equation}
\frac{dP}{dR_e} = \frac{\partial P}{\partial R_e} + \sum_j \frac{\partial P}{\partial I_j} \cdot \frac{d I_j}{d R_e}
= I_e^2 + \sum_j 2 R_j I_j \cdot \frac{d I_j}{d R_e}.
\end{equation}

The second term vanishes because \( \mathbf{I}[\mathbf{R}] \) satisfies the constraint \( A \mathbf{I} = \mathbf{s} \) for all \( \mathbf{R} \), so \( A \cdot \frac{d\mathbf{I}}{dR_e} = 0 \). Moreover, the partial derivative of \( P \) with respect to \( I_j \) is \( \frac{\partial P}{\partial I_j} = 2 R_j I_j = - (A^\top \boldsymbol{\lambda})_j \) by the stationarity condition. Hence

\begin{equation}
\sum_j \frac{\partial P}{\partial I_j} \cdot \frac{d I_j}{d R_e} = - \sum_j (A^\top \boldsymbol{\lambda})_j \cdot \frac{d I_j}{d R_e}
= -\boldsymbol{\lambda}^\top A \cdot \frac{d \mathbf{I}}{d R_e} = 0.
\end{equation}

Therefore,

\begin{equation}
\frac{dP}{dR_e} = I_e^2.
\end{equation}

This shows that even though \( I_e \) depends implicitly on \( R_e \), the total derivative of the power dissipation with respect to \( R_e \) is simply the local squared current \( I_e^2 \).

Now suppose resistances evolve to minimize dissipation subject to a cost constraint of the form \( C[\mathbf{R}] = \sum_e c(R_e) \). Define the total potential

\begin{equation}
\Phi[\mathbf{R}] = P[\mathbf{R}] + \lambda C[\mathbf{R}] = \sum_e R_e I_e^2 + \lambda \sum_e c(R_e).
\end{equation}

A gradient descent evolution of \( R_e \) to reduce \( \Phi \) yields

\begin{equation}
\frac{d R_e}{d t} = - \frac{d\Phi}{dR_e} = - I_e^2 - \lambda \frac{dc}{dR_e}.
\end{equation}

Defining \( \tilde{\lambda} = -\lambda \), we obtain the local resistance adaptation rule

\begin{equation}
\boxed{
\frac{d R_e}{d t} = -I_e^2 + \tilde{\lambda} \cdot \frac{d c}{d R_e}
}
\end{equation}

This equation shows that resistance evolves locally: the change in \( R_e \) depends only on the local current \( I_e \) and the local marginal cost \( \frac{d c}{d R_e} \). It captures how global dissipation minimization under current conservation and cost constraints can be implemented by local, distributed rules.

We may re-express this rule in terms of the local dissipation \( P_e = R_e I_e^2 \). Multiplying both sides by \( R_e \) gives

\begin{equation}
R_e \cdot \frac{d R_e}{dt} = - R_e I_e^2 + \tilde{\lambda} \cdot R_e \cdot \frac{d c}{d R_e} = -P_e + \tilde{\lambda} R_e \cdot \frac{dc}{dR_e}.
\end{equation}

This form emphasizes that resistance decreases proportionally to the dissipation on the edge, offset by a marginal cost pressure. It also reveals that the square of the resistance evolves according to

\begin{equation}
\boxed{
\frac{d(R_e^2)}{dt} = -2 P_e + 2 \tilde{\lambda} R_e \cdot \frac{dc}{dR_e},
}
\end{equation}

highlighting a natural edge-local interpretation of the adaptation dynamics in terms of reinforcing local dissipative flows.

\section{Complexity decreases with the circuit size} \label{satur_complexity}
\begin{figure}[ht!]
    \centering
    \includegraphics[width=0.8\linewidth]{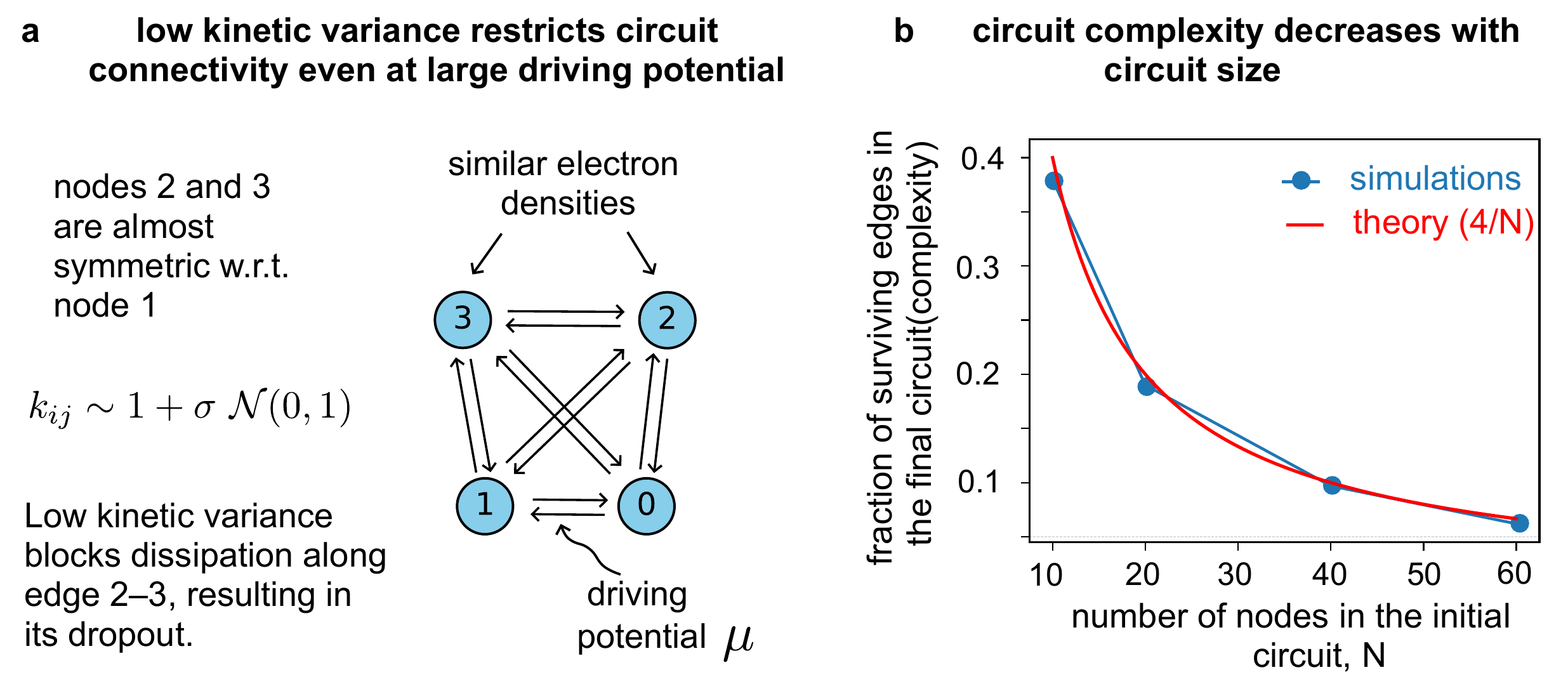}
    \vspace{0.5cm}
    \caption{\justifying \textsf{\textbf{Complexity decreases with circuit size} (a) The survival of an electron transfer pathway between two nodes—defined by whether its edge dissipation exceeds the maintenance threshold—depends on their relative electron affinities within the broader circuit, especially in relation to the energy-coupled (battery) edge and its adjacent nodes. When abiotic (static) transition rates exhibit low variance, dissipation along edges distant from the energy source tends to fall below maintenance, causing those connections to collapse. As a result, at sufficiently high driving potentials, the circuit converges to a near-deterministic topology. (b) A simple combinatorial estimate ($\sim 4/\text{N}$, refer text) for the fraction of surviving edges in low-variance, single-battery living circuits accurately captures the saturation behavior observed in numerical simulations of circuit complexity.}}
\end{figure}

A sufficiently large driving potential can break detailed balance, leading to non-equilibrium steady states where only a subset of edges in the original circuit remains active. As discussed in the main text, the number of surviving edges increases with driving strength before saturating at a fixed value. For an $N$-node circuit driven by a single energy source and with low variance in kinetic parameters---i.e., $k_{ij} \sim 1 + \sigma \eta_{ij}$ where $\eta_{ij}\sim\mathcal{N}(0,1)$--- for $\sigma \ll 1$, we can argue how the fraction of surviving edges approaches approximately $4/N$.

To illustrate this, consider a 4-node circuit with the $0$--$1$ edge coupled to a driving source (Fig. S2a). Due to the near-uniformity of the kinetic rates, the steady-state densities at nodes $2$ and $3$ are nearly equal, resulting in minimal flux across the $2$--$3$ edge. Since dissipation depends on flux, such edges fail to surpass the dissipation threshold $\sigma_{\text{maint}}$ and collapse. Only edges directly connected to the source nodes ($0$ and $1$), along with the source edge itself, remain active. In total, there are $2(N-2)$ such edges plus the driven edge, yielding:
\[
\frac{2(N - 2) + 1}{\binom{N}{2}} = \frac{2(2(N-2)+1)}{N(N-1)} \approx \frac{4}{N}
\]

\section{Drive-complexity relationship for multiple sources of nonequilibrium drive}
\begin{figure}[ht!]
\centering
\includegraphics[width=0.5\linewidth]{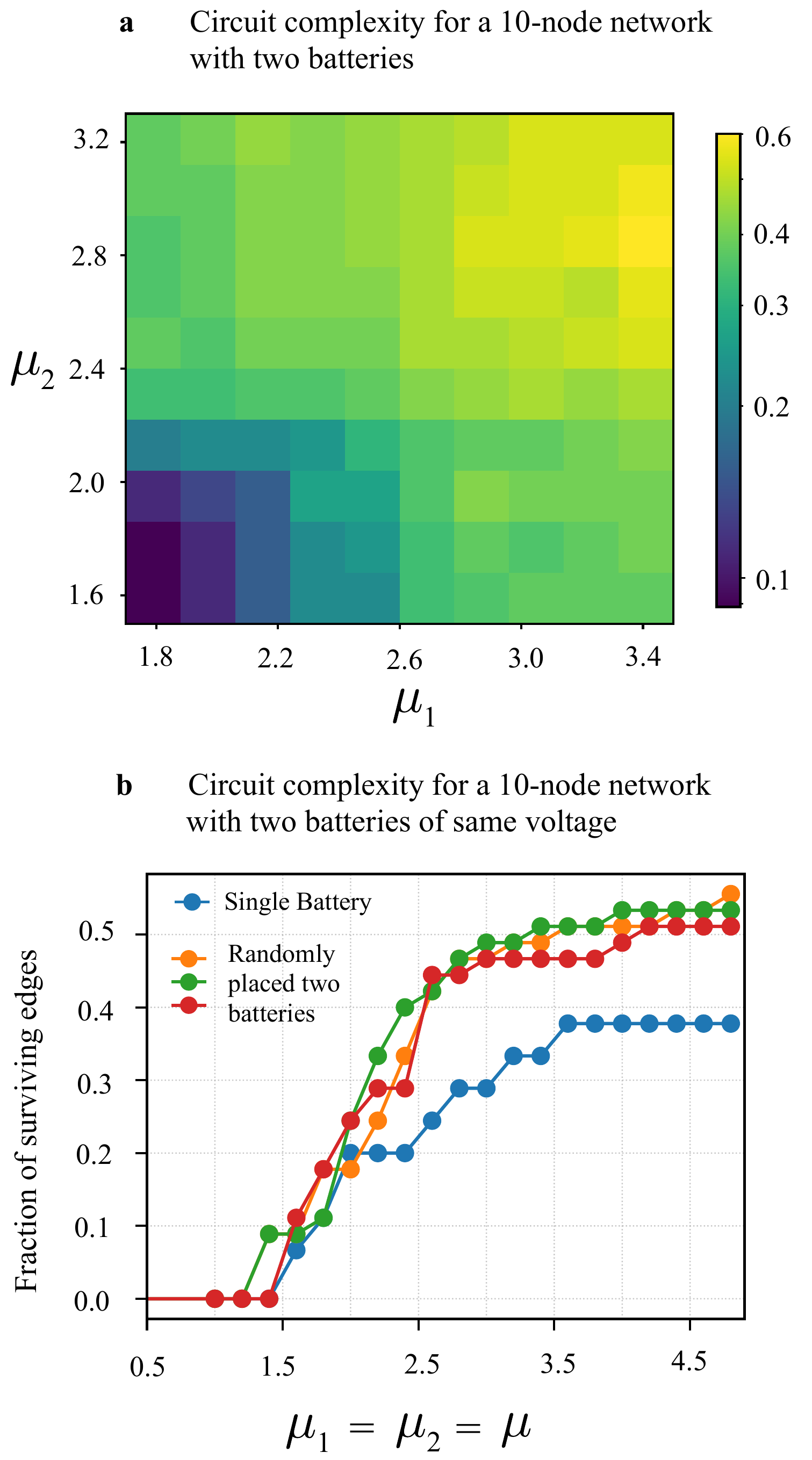} \vspace{0.1cm}
 \caption{\justifying \textsf{\textbf{Multiple batteries enable the emergence of more complex networks.} (a) A 10-node circuit with two edges connected to separate external energy sources can support a greater number of edges, promoting the development of more complex structures. The heatmap shows the fraction of surviving edges in the final self-organized non-equilibrium steady state, out of the initial 45 edges in the 10-node circuit. (b) To assess the role of battery placement, two edges are randomly selected and coupled to batteries of equal voltage. Regardless of their relative positions, the introduction of multiple batteries consistently enhances circuit complexity.}}

    \label{fig:multBattFig}
\end{figure}
In living circuits, edges corresponding to autotrophs play the central role in driving the system toward self-organization. Two key factors— the initial network topology and the relative positions of the autotrophic edge(s)— strongly influence which of the original edges persist in the final self-organized configuration. As discussed in the main text, in a 4-node network, the relative positioning of two autotrophs determines whether they compete or cooperate along a shared edge. We now extend this analysis to 10-node networks, where two edges are randomly selected and coupled to external energy sources. We observe that, in general, the presence of two autotrophs (or batteries) consistently promotes greater circuit complexity compared to configurations with a single energy source (see Fig. \ref{fig:multBattFig}).

\end{document}